\documentclass[prb,singlecolumn,showpacs,preprintnumbers,amsmath,amssymb]{revtex4}

\usepackage{amsfonts}
\usepackage{latexsym}
\usepackage{graphicx}
\usepackage[usenames]{color}
\newcommand{\Fref}[1]{Fig.~\ref{#1}}
\newcommand{\Eqref}[1]{Eq.~(\ref{#1})}

\newcommand{\nn}{\nonumber}
\newcommand{\be}{\begin{equation}}
\newcommand{\ee}{\end{equation}}

\newcommand{\bear}{\begin{eqnarray}}
\newcommand{\ear}{\end{eqnarray}}

\newcommand{\ii}{\mathrm{i}}

\begin{document}

\title{Amplification of Slow Magnetosonic Waves by Shear Flow:\\
Heating and Friction Mechanisms of Accretion Disks
   \footnote
   {in \emph{Space Plasma Physics,}
   Proceedings of the School and Workshop on
   Space Plasma Physics,
   31 August--7 September 2008,
   Sozopol, Bulgaria, Editor: I.~Zhelyazkov, American Institute of Physics,
   AIP Conference Proceedings \textbf{1121}, pp.~28-54 (2009),
   ISBN-978-0-7354-0-0656-8, ASSN-0094-243X.}
}

\author{T.M.~Mishonov}
\email[E-mail: ]{tmishonov@phys.uni-sofia.bg}
\author{Z.D.~Dimitrov}
\email[E-mail: ]{Zlatan.Dimitrov@gmail.com}
\author{Y.G.~Maneva}
\email[E-mail: ]{yanamaneva@gmail.com}
\author{T.S.~Hristov}
\email[E-mail: ]{Tihomir.Hristov@jhu.edu}
\affiliation{Department of Theoretical Physics, Faculty of Physics,\\
University of Sofia ``St.~Clement of Ohrid'',\\
5 J. Bourchier Blvd, BG-1164 Sofia, Bulgaria}


\date{April 9, 2009}

\begin{abstract}
Propagation of three dimensional magnetosonic waves is considered
for a homogeneous shear flow of an incompressible fluid.  The
analytical solutions for all magnetohydrodynamic variables are
presented by confluent Heun functions.  The problem is reduced to
finding a solution of an effective Schr\"odinger equation.  The
amplification of slow magnetosonic waves is analyzed in great
details.  A simple formula for the amplification coefficient is
derived. The velocity shear primarily affects the incompressible
limit of slow magnetosonic waves.  The amplification is very strong
for slow magnetosonic waves in the long-wavelength limit. It is
demonstrated that the amplification of those waves leads to
amplification of turbulence. The phenomenology of Shakura--Sunyaev
for the friction in accretion disks is derived in the framework of
the Kolmogorov turbulence.  The presented findings may be the key to
explaining the anomalous plasma heating responsible for the
luminosity of quasars.  It is suggested that wave amplification is
the keystone of the self-sustained turbulence in accretion disks.

\end{abstract}

\maketitle

\section{Introduction}                                  %
%
The occurrence of intense dissipation in accretion flows is among
the long-standing unsolved problems in
astrophysics.\cite{Balbus:91,Fridman:08} Revealing how the
magnetized turbulence creates shear stress tensor is of primary
importance to understand the heating mechanism and the transport of
angular momentum in accretion disks. The transport of angular
momentum at greatly enhanced rates is important for the main problem
of cosmogony, that is understanding the dynamics of creation of
compact astrophysical objects.\cite{Balbus:98,Balbus:03} Without a
theory explaining the enhanced energy dissipation in accretion flows
of turbulent magnetized plasma we would have no clear picture of how
our solar system has been created, why the angular momentum of the
Sun is only $2$\% of the angular momentum of solar system, while
carrying 99\% of the solar system's mass, why quasars are the most
luminous sources in the universe. The importance of friction forces
and convection as well the problem of angular momentum
redistribution for the first time was emphasized by von
Weitz\"acker;\cite{Weitzaecker:48} a very detailed bibliography on
the physics of disks is provided in the monograph by Morozov and
Khoperskov.\cite{Morozov:05} Gravitational forces, angular momentum
conservation, and dissipation processes become the main ingredients
of the standard model of accretion disks by Shakura and
Sunyaev\cite{Shakura:73} and Lynden-Bell and
Pringle.\cite{Lynden-Bell:74} Lynden-Bell\cite{Lynden-Bell:69}
suggested that quasars are accretion disks and Shakura and Sunyaev
introduced alpha phenomenology for the stress tensor
\be \sigma_{R\varphi}=\alpha p,\qquad p\sim\rho c_\mathrm{s}^2, \ee
where $\alpha$ is a dimensionless parameter, $p$ is the pressure
$\rho$ is the mass density, $c_\mathrm{s}$ is the sound speed,
and the indices of the tensor come from the cylindrical
coordinate system $(R,\varphi,z)$ related to the disk rotating
around the $z$-axis. As accretion disks can have completely
different scales for protostellar disks, mass transfer disks, and
disks in active galactic nuclei (AGN) it is unlikely that Coriolis
force is the main cause for dissipation, while the bending of the
trajectories is a critical ingredient.  We suppose that the
shear dissipation in magnetized turbulent plasma is a robust and
very general phenomenon which can be analyzed as a local heating for
approximately homogeneous magnetic field and gradient of the
velocity. For these reasons in the next section we will use local
Cartesian coordinates choosing the $x$-axis in radial direction
$\mathbf{e}_x\equiv\hat R,$ and the $y$-axis along the circulation of
the almost Keplerian motion of accreting plasma,
$\mathbf{e}_y\equiv\hat \varphi.$  The axis of the disk is along
$\mathbf{e}_z\equiv \hat z$.

There is almost a consensus that the magnetic field is essential and
should be introduced from the very beginning in the
magnetohydrodynamic (MHD) analysis. The likely importance of MHD
waves on the cosmogony of solar system has been pointed out by
Alfv\'en: ``At last some remarks are made about the transfer of
momentum from the Sun to the planets, which is fundamental to the
theory. The importance of the magnetohydrodynamic waves in this
respect is pointed out.\cite{Alfven:46}'' As the magnetic field
lines are frozen in the highly conducting hot plasma, we suppose
that they lie in the plane of the disk $\mathbf{B}\approx -B
\mathbf{e}_\varphi$.  In our local analysis we can suppose that mass
density is also constant $\rho\approx\mathrm{const}$.  Moreover, we
assume that the magnetic field is sufficiently small and the
Alfv\'en speed is much smaller than the sound speed, $V_\mathrm A\ll
c_\mathrm{s}$.  As usual, the magnetic pressure is defined as
\be
\frac{1}{2}\rho V_\mathrm A^2=p_B\equiv\frac{B^2}{2\mu_0}.
\ee
Under these conditions, we can consider the plasma as an
incompressible fluid. In other words we assume a plasma for
which the magnetic pressure $p_B$ is much smaller than the gaseous
one $p$
\be
\beta\equiv\frac{p}{p_B}\gg 1.
\ee
According to the theory of turbulent dynamo the density of magnetic
energy $p_B$ is comparable with the density of kinetic energy of the
turbulent motion at a basic scale which is smaller than the pressure,
i.e., we suppose that the turbulence creating the magnetic field is
subsonic.

Our goal is to demonstrate that small magnetic fields can catalyze a
disk to ignite as a star and to help the kinetic energy of the shear
flow to be converted into heat in spite of the very low molecular
plasma viscosity $\eta$.

In a weak magnetic field the incompressible transverse MHD waves,
the slow magnetosonic waves (SMWs) and Alfv\'en waves (AWs), are
qualitatively new features of the weakly magnetized plasma.  Under
these conditions every mechanism of giant wave amplification
inevitably acts as a dissipation mechanism for heat
production which creates stress tensor and a transport of
angular momentum in accretion disks.

The present research is triggered by the amplification of SMWs
observed by numerical analysis of two-dimensional (2D) MHD
waves.\cite{Chagelishvili:93} Even these first investigations
demonstrated that shear flow leads to exchange of energy and wave
amplification, mutual transformation between different wave modes,
gave the perspectives to explain self-heating and other processes
now known as \textit{nonmodal.} As we have standing wave
amplification often is used the notion \emph{overreflection}; a
running wave with $k_y$ is amplified together with the reflected
wave with wave-vector $-k_y$. These conclusions was confirmed
extended and popularized by the numerical
investigations.\cite{Rogava:03} However without a realistic
analytical three-dimensional solution the investigations of shear
flows is in state of infancy.  The purpose of our work is to
analytically solve the simplest case of three-dimensional (3D) waves
in homogeneous shear flow and magnetic field, to analytically
calculate the phase averaged amplification coefficient $\mathcal{A}$
of the waves and to incorporate this waves's amplification into some
standard model for turbulence. Or, in short, our final aim is to use
an exact solution for linearized waves in order to build up a
coherent scenario for accretion disk theory and luminosity of
quasars.  Before continuing we wish to refer to the concluding
remarks from the review by Balbus and Hawley:\cite{Balbus:98}
\textit{Again and again we are ignorant. The good news is that, for
first time, it appears that we know in which directions we should be
looking to begin to find answers to questions like these.}

\section{MHD model. Analytical Solution to Linearized Wave Equations} %

Our model problem is to investigate MHD waves in a homogeneous
shear flow and a homogeneous magnetic field
\begin{eqnarray}
&&\mathbf{V}_\mathrm{shear}= Ax\mathbf{e}_y,\qquad \mathbf{B}_0
=-B_0\mathbf{e}_y,\qquad \mathbf{e}_i=\partial_i \mathbf{r},
\qquad \mathbf{e}_B\equiv\mathbf{B}_0/B_0=-\mathbf{e}_y,\\
&&\bm{\Omega}=\nabla\times\mathbf{V}_\mathrm{shear} = A\mathbf{e}_z,\qquad
i=1,2,3,\qquad \mathbf{e}_\Omega=\bm{\Omega}/\Omega= \mathbf{e}_z
\end{eqnarray}
in an ideal and incompressible plasma. For the linearized MHD equations
we suppose a time dependent wave-vector
\begin{equation}
\label{k(t)}
\mathbf{k}(t)=-Atk_y\mathbf{e}_x+k_y\mathbf{e}_y+k_z\mathbf{e}_z,
\qquad \mathbf{k}\cdot\mathbf{r}=(y-Atx)k_y + k_z z
\end{equation}
in the plane waves solution.\cite{Townsend:76} For the velocity
$\mathbf{V}$ and the magnetic field $\mathbf{B}$ we consider small
enough time-dependent amplitudes $\mathbf{v}$ and $\mathbf{b}$
\begin{eqnarray}
\label{wave}
&\mathbf{V}=\mathbf{V}_\mathrm{shear}+\mathbf{V}_\mathrm{wave}\, ,\qquad
&\mathbf{B}=\mathbf{B}_0+\mathbf{B}_\mathrm{wave}\, ,
\\
&\mathbf{V}_\mathrm{wave}=-
\ii V_\mathrm{A}\mathbf{v}(t)\mathrm{e}^{\ii\mathbf{k}\cdot\mathbf{r}},
&\mathbf{B}_\mathrm{wave}
=B_0\mathbf{b}(t)\mathrm{e}^{\ii\mathbf{k}\cdot\mathbf{r}}.\nn
\end{eqnarray}
The magnetic field $\mathbf{B}$ and velocity $\mathbf{V}$ are, of
course, real vector fields as functions of $t$ and the position vector
$\mathbf{r}=(x,y,z)$, but it is mathematically convenient to use
complex variables and calculate the real parts of the derived
solutions.

Let us summarize the supposed approximations: incompressible
$\nabla\cdot\mathbf{V}=0,$ and inviscid $\eta=0,$ fluid with
constant density $\rho=\mbox{const}$, and negligible Ohmic
resistivity $\varrho=0.$ For a sufficiently slow nonrelativistic
motion the MHD equations take the form\cite{LL8}
\be
\label{MHD}
\rho \mathrm{D}_t \mathbf{V}
= - \nabla p -\mathbf{B}\times (\nabla\times\mathbf{B})/\mu_0,
\qquad
\mathrm{D}_t\mathbf{B}= \mathbf{B}\cdot\nabla \mathbf{V},
\ee
where $\mathrm{D}_t \equiv \partial_t + \mathbf{V}\cdot\nabla$ is the
substantial Lagrange derivative; $\nabla\cdot\mathbf{B}=0$.

Let us introduce the characteristic length-scale $\Lambda$ and velocity
$V_{\mathrm A}$ of the problem
\be
\label{Lambda}
\Lambda\equiv \frac{V_\mathrm A}{A}\ll R,\qquad
\mathbf{V}_\mathrm A\equiv\frac{\mathbf{B}_0}{\sqrt{\mu_0\rho}},\qquad
V_\mathrm A\equiv |\mathbf{V}_\mathrm A| \ll c_\mathrm{s},
\ee
dimensionless wave-vector $\mathbf{K}\equiv\Lambda \mathbf{k}$, and
dimensionless time
\be
\tau\equiv At=-\frac{K_x(t)}{K_y}=-\frac{k_x(t)}{k_y}.
\ee
The initial time is chosen so that $k_x(t=0)=0$.
We suppose that the characteristic length $\Lambda$ is much smaller
than the disk thickness $d_\mathrm{disk}$.

As the shear parameter describes the radial gradient of the orbiting
velocity
\be
A\equiv \frac{\partial V_\varphi}{\partial R}
=-\frac{1}{2}\omega_\mathrm{Kepler}
\ee
at the corresponding radius $R$, it is of order of the angular
frequency $\omega_\mathrm{Kepler}$ of the Keplerian motion. We
suppose also that for weak magnetic fields the Alfv\'en speed is
much smaller than the velocity of Keplerian orbiting, $V_\mathrm
A\ll V_\mathrm{Kepler}\equiv R\omega_\mathrm{Kepler}$.  In this case
the wavelength of the slow magnetosonic waves is much smaller than the
radius of the orbit $R\sqrt{k_\varphi^2+k_z^2}\gg 1$ and the disk
thickness $|k_z|d_\mathrm{disk}\gg1$ and we approximately consider
the shear flow as homogeneous. In other words, the goal of the
present research is to investigate the statistical properties of a
shear flow determined by short wavelength waves inaccessible by
direct astronomical observations and usual computer simulations of
accretion disk magnetohydrodynamics.

The Coriolis force has a negligible influence on the propagation of
magnetohydrodynamic waves and their amplification analyzed in the
present work for the case of dominating magnetic field.  Consequently,
in our further consideration we will neglect that force. In
such a way we have to investigate the influence of a homogeneous
shear flow on small amplitude MHD waves in an incompressible ideal
fluid. The homogeneous magnetic field we suppose to be parallel to
the shear flow.

With so introduced variables the substitution of wave components
from \Eqref{wave} in MHD equations \Eqref{MHD} after some algebra
yields\cite{Chagelishvili:93}
\begin{eqnarray}
\label{Tihomir93Eqn}
\label{x}
 &&\mathrm{d}_{\tau} v_x = 2\frac{K_yK_x}{K^2}v_x - K_y b_x, \qquad
 \mathrm{d}_{\tau} b_x = K_y v_x,
\\
\label{z}
 &&\mathrm{d}_{\tau} v_z = 2\frac{K_yK_z}{K^2}v_x - K_y b_z, \qquad
 \mathrm{d}_{\tau} b_z = K_y v_z,
\\
\label{div0}
&&\mathbf{K}\cdot\mathbf{v}=0=
\mathbf{K}\cdot\mathbf{b}, \qquad
\mathbf{K}=-K_y\tau\mathbf{e}_x+K_y\mathbf{e}_y+K_z\mathbf{e}_z.
\end{eqnarray}
An unabridged derivation of this set of equations is given in
Ref.~[\onlinecite{Mishonov:07}]. Later on, the substitution of
$\mathrm{d}_{\tau} v_x$ with $\mathrm{d}_{\tau}^2 b_x$ in \Eqref{x},
and analogous substitution of $\mathrm{d}_{\tau} v_z$ in \Eqref{z}
leads to
\begin{eqnarray}
\label{bx}
&&\mathrm{d}_{\xi}^2b_x + \frac{2\xi}{1 + \xi^2}
\mathrm{d}_{\xi}b_x + Q^2b_x=0, \\
&&\label{bz}
\mathrm{d}_{\xi}^2b_z + Q^2b_z = 2K_z
\frac{v_x}{1+\xi^2},
\end{eqnarray}
where
\begin{eqnarray}
&&\xi\equiv\tau\frac{K_y}{Q}=-\frac{K_x}{\sqrt{K_y^2+K_z^2}},\\
&&Q\equiv\sqrt{K_y^2+K_z^2}=\mathrm{const.}
\end{eqnarray}
For a two-dimensional motion ($K_z=0$) in the plane of the disk $\xi=\tau$.
The negative friction for $\xi<0$ in the oscillator equation \Eqref{bx}
comes from the time-dependence of the $x$-component of the wave-vector
\be
K_x(\tau)=-K_y \tau=-\,Q\xi.
\ee
The other components are constant
$K_y=\mathrm{const}$, and $K_z=\mathrm{const}$. In such a way the
change of the sign in the effective friction force is a property of
the shear flow related to the change of the sign of the corresponding
component of the wave-vector.

For the Alfv\'en wave amplitude $b_z$ in \Eqref{bz} we have the
equation of a harmonic oscillator with an external force. This equation
has the Green function
\begin{eqnarray}
 && \left( \mathrm{d}_{\xi}^2 + Q^2 \right) G(\xi, \xi^\prime)
 =\delta(\xi-\xi^\prime)=\mathrm{d}_\xi\theta(\xi-\xi^\prime),\\
 && G(\xi, \xi^\prime)= \frac{\sin[Q(\xi-\xi^{\prime})]}{Q}\theta(\xi-\xi^\prime),
\end{eqnarray}
and the general solution to \Eqref{bz} reads
\begin{eqnarray}
\label{bzfin}
&&b_z(\xi)=
\frac{2K_z}{Q} \int_{-\infty}^{\xi} \sin[Q(\xi-\xi^{\prime})]
\frac{v_x(\xi^{\prime})}{1+\xi^{\prime 2}}\mathrm{d}\xi^{\prime}+\chi(\xi),
\nn
\\
\label{chi_general}
&&
\left(\mathrm{d}_\xi^2+Q^2\right)\chi(\xi)=0,\qquad
\chi(\xi)=\tilde{C}_\mathrm g \chi_\mathrm{g}
+\tilde{C}_\mathrm u \chi_\mathrm{u},
\nn\\
&&\chi_\mathrm{g}(\xi)=\cos(Q\xi),\qquad
\chi_\mathrm{u}(\xi)=\frac{\sin(Q\xi)}{Q},
\end{eqnarray}
where $\tilde{C}_\mathrm g$ and $\tilde{C}_\mathrm u$ are the
amplitudes the even and the odd eigen $\chi$-oscillations related to
the pure Alfv\'en waves (AWs) for the case of $K_z=0$. AWs are not
influenced by the shear flow. For the solution of the Cauchy problem
with fixed $b_z(\xi_0)$ and $\mathrm{d}_\xi b_z(\xi_0)=Qv_z(0)$ we have
\be
b_z(\xi)=
\frac{2K_z}{Q} \int_{\xi_0}^{\xi} \sin[Q(\xi-\xi^{\prime})]
\frac{v_x(\xi^{\prime})}{1+\xi^{\prime 2}}\mathrm{d}\xi^{\prime}
+b_z(\xi_0)\cos[Q(\xi-\xi_0)]
+\mathrm{d}_\xi b_z(\xi_0)\frac{\sin[Q(\xi-\xi_0)]}{Q}.
\label{Green_Cauchy}
\ee

For the special case of $K_z=0$ we have a complete separation of
variables in two independent sets of equations.  One of them is for
Alfv\'en waves  with oscillating $b_z$ and $v_z$ perpendicular
to the wave-vector and magnetic field, and the second set is for the
independent slow magnetosonic waves (SMWs) for which the
oscillations of the magnetic field and velocity are in the plane of
the external magnetic field and the wave-vector.  Still we can use
the terminology AWs and SMWs for waves having nonzero $K_z$.  In
this case the amplitude of the SMW $b_x$ plays the role of external
driving force for the amplitude of AW in \Eqref{bz}. We will see
later that amplification is related to SMWs, and AWs with $b_z$
components are only influenced by SMWs at nonzero $K_z$.  Due to a
resonance this influence, however, could be significant. It is
notable that AW amplitude $b_z$ has provides no feedback on the SMW
amplitude $b_x$.

MHD equations give the possibility to express the velocity by the
magnetic field. For the velocity from \Eqref{x} and \Eqref{z} we have
\be v_z(\xi)=\frac{\mathrm{d}_\xi b_z(\xi)}{Q},\qquad
v_x(\xi)=\frac{\mathrm{d}_\xi b_x(\xi)}{Q},\qquad
\mathrm{d}_\xi \equiv \frac{Q}{K_y}\mathrm{d}_\tau.
\ee
The incompressibility conditions \Eqref{div0} yield explicit
expressions for the oscillations parallel to the external magnetic
field and shear flow
\be
\label{byfin}
b_y=-\frac{K_xb_x+K_zb_z}{K_y}
   =-\frac{-K_y\tau b_x+K_zb_z}{K_y}
   =\frac{Q}{K_y}\xi b_x-\frac{K_x}{K_y}b_z
, \qquad
v_y=-\frac{K_xv_x+K_zv_z}{K_y}=\frac{Q}{K_y}\xi v_x-\frac{K_x}{K_y}v_z.
\ee

In such a way the main detail in solving the set of \Eqref{x} and
\Eqref{z} is to derive an analytic formula for $b_x$. The first
derivative $\mathrm{d}_{\xi}b_x$ term in \Eqref{bx} disappears by
the substitution
\be
\label{bxfin}
b_x(\xi)=\frac{\psi(\xi)}{\sqrt{1+\xi^2}},
\ee
and we arrive at an effective Schr\"odinger equation\cite{Rogava:03}
which describes the SMW
\be
\label{Sch}
\mathrm{d}_\xi^2\psi+\left[Q^2-\frac{1}{(1+\xi^2)^2}\right]\psi=0.
\ee
This ``Schr\"odingerization'' gives the possibility to use the
quantum mechanical analogies\cite{Landau:32} and also the whole
system of notions of quantum scattering theory. Constant Wronskians
give also a technical convenience. As we will see later the
amplification of SMW in 2D case ($k_z=0$) is represented by the
asymptotics phases of the wavefunctions \Eqref{ampl-trans}; a simple
illustration of Heisenberg 1938 idea that S-matrix contains whole
experimentally accessible information.

As the effective potential $2m U/\hbar^2=1/(1+\xi^2)^2$ is an even
function of $\xi$, the general solution to the Schr\"odinger equation
\Eqref{Sch} is a linear combination of the even and odd solutions
\begin{eqnarray}
\label{gen}
&&\psi(\xi)=C_\mathrm g \psi_\mathrm g(\xi) + C_\mathrm u \psi_\mathrm u(\xi),\\
&&\psi_\mathrm g(-\xi)=\psi_\mathrm g(\xi),
\qquad \psi_\mathrm u(-\xi)=-\psi_\mathrm u(\xi),\nn
\end{eqnarray}
which obey the boundary conditions
\begin{eqnarray}
&&\psi_\mathrm{g}(0)=1,\qquad \mathrm{d}_\xi\psi_\mathrm{g}(0)=0,\\
&&\psi_\mathrm{u}(0)=0,\qquad \mathrm{d}_\xi\psi_\mathrm{u}(0)=1.
\end{eqnarray}
The analytical solutions are represented by the confluent Heun function
\begin{eqnarray}
\label{PsiHeun_g}
\psi_\mathrm g\!\!=\!\sqrt{1+\xi^2}
\,\mathrm{HeunC}(0,-\frac{1}{2},0,-\frac{Q^2}{4},\!\frac{1+Q^2}{4},-\xi^2),\\
\label{PsiHeun_u}
\psi_\mathrm u\!\!=\!\xi\sqrt{1+\xi^2}
\,\mathrm{HeunC}(0,+\frac{1}{2},0,-\frac{Q^2}{4},\!\frac{1+Q^2}{4},-\xi^2).
\end{eqnarray}
In the computer algebra system Maple the confluent Heun function reference is
$\mathrm{HeunC}(\alpha,\beta,\gamma,\delta,\eta,z)$.  This special
function obeys the equation
\be
z(z-1)y^{\prime\prime}
+ [\mathsf{A}z^2+\mathsf{B}z+\mathsf{C}]\,y'
+[\mathsf{D}z+\mathsf{E}]\,y=0,
\ee
where
\begin{eqnarray}
&&\mathsf{A}=\alpha,\qquad
\mathsf{B}=2+\beta+\gamma-\alpha, \qquad
\mathsf{C}=-1-\beta,\qquad
\\
&& \mathsf{D}=\frac{1}{2}[(2+\gamma+\beta)\alpha+2\delta],\qquad
\mathsf{E}=\frac{1}{2}[-\alpha(1+\beta)+(1+\gamma)\beta + \gamma +2\eta)].
\end{eqnarray}
For the series expansion
\be
y=\sum_{n=0}^\infty c_n =\sum_{n=0}^\infty a_n z^n,\qquad c_n\equiv a_nz^n
\ee
we arrive at the recursion
\be
\mathtt{F}a_{n-1}+\mathtt{G}a_n+\mathtt{H}a_{n+1}=0,
\ee
where
\begin{eqnarray}
&&\mathtt{F}=\delta +\frac{1}{2}(2n+\beta\gamma),\\
&&\mathtt{G}=n^2+(1+\gamma+\beta-\alpha)n
+\frac{1}{2}[\gamma+2\eta-\alpha +\beta(1-\alpha+\gamma)]\nn\\
&&\mathtt{H}=-(n+1)(n+1+\beta).
\end{eqnarray}
Supposing $a_{-1}=0$ and $a_0=1$ we use the recursion
\be
a_{n+1}=-(\mathtt{F}a_{n-1}+\mathtt{G}a_n)/\mathsf{H}
\ee
which, for example, gives
\be
a_1=\frac{1}{2}\left[\beta(1+\gamma-\alpha)+\gamma+2\eta-\alpha\right]/(1+\beta).
\ee
As the effective Schr\"odinger equation \Eqref{Sch} has a solution for
arbitrary $\xi\in(-\infty,\;+\infty)$, the formal series for Heun
function have convergent Pad\'e approximants.

Those Pad\'e approximants can be calculated by the well-known
$\varepsilon$-algorithm. First we calculate the series of the
partial sums in zeroth $(0)$ approximation
\be
S_i^{(0)}=S_{i-1}^{(0)}+c_i
\ee
and for the first $N$ terms we get
\begin{eqnarray}
&&S_0^{(0)}=c_0, \qquad S_1^{(0)}= c_0+c_1,\qquad\dots,\qquad
S_N^{(0)}=c_0+c_1+\dots +c_N.
\end{eqnarray}
Our problem is to calculate the limit of the sequence
\be
S=\lim_{n\rightarrow\infty} S_n.
\ee
For this calculation we generate the auxiliary sequence
\be
H_i^{(0)}=1/c_{i+1}
\ee
recalling that $1/0=0$, i.e., using pseudoinverse numbers if we have
to divide by zero:
\be
H_0^{(0)}=1/c_{1},\qquad
H_1^{(0)}=1/c_{2},
\qquad \dots, \qquad
H_{N-1}^{(0)}=1/c_{N}.
\ee

The epsilon-algorithm is the calculation of the recursion for
the series
\begin{eqnarray}
&&S_i^{(k)}=S_{i+1}^{(k-1)}+1/\left(H_{i+1}^{(k-1)}-H_i^{(k-1)}\right),\\
&&H_i^{(k)}=H_{i+1}^{(k-1)}+1/\left(S_{i+1}^{(k-1)}-S_i^{(k-1)}\right)
\end{eqnarray}
for all indices for which these relations make sense.

The maximal in modulus auxiliary element
$H_\mathrm{max}=\left|H_{I+1}^{(K)}\right|$ gives the best Pad\'e
approximant $S_{I}^{(K)}$ for the searched limes and the accuracy is
of the order of $1/H_\mathrm{max}$.

For each accuracy of the final result $\epsilon$ we can calculate
$S$ and $H$ sequences with an accuracy $\delta$ to assure that
$1/H_\mathrm{max}< \epsilon$.  In such a way we obtain a method for
calculating the confluent Heun function and the solution to the MHD
equation in power series of time. If from physical arguments we know that a
solution exists, the divergent series can be
summed. Having a method for calculating the confluent Heun
function, we can calculate both the even and odd solutions to
\Eqref{gen}. The accuracy in calculating Heun functions is
controlled by the Wronskian
\be
W(\psi_\mathrm{g},\psi_\mathrm{u})
=
\left|\begin{array}{cc}
\psi_\mathrm{g}(\xi)&\psi_\mathrm{u}(\xi)\\
\mathrm{d}_\xi\psi_\mathrm{g}(\xi)&\mathrm{d}_\xi\psi_\mathrm{u}(\xi)
\end{array}\right|
=1.
\ee
The constants from the general solution are also given by the
Wronskians
\be
C_\mathrm{g}=W(\psi,\psi_\mathrm{u})
=\psi(\xi_0)\mathrm{d}_\xi\psi_\mathrm{u}(\xi_0)
-\psi_\mathrm{u}(\xi_0)\mathrm{d}_\xi\psi(\xi_0)
,\qquad
C_\mathrm{u}=W(\psi_\mathrm{g},\psi)
=\psi_\mathrm{g}(\xi_0)\mathrm{d}_\xi\psi(\xi_0)
-\psi(\xi_0)\mathrm{d}_\xi\psi_\mathrm{g}(\xi_0)
.
\label{CoefficientsByWronskians}
\ee
Those formulae generally apply for a Cauchi problem where the
initial conditions are imposed on the function being
sought, i.e., on $\psi(\xi_0)$ and its derivative
$\mathrm{d}_\xi\psi_\mathrm{g}(\xi_0).$

All MHD variables can be expressed by the solution \Eqref{gen} to
the effective Schr\"odinger equation. Then $b_x(\xi)$ is given by
\Eqref{bxfin}, and $b_z(\xi)$ by \Eqref{bzfin}. Equation
(\ref{byfin}) gives $b_y$.  And finally we know the time
$\tau=Q\xi/K_y$ dependent amplitude of the magnetic field
$\mathbf{b}(\xi=K_y\tau/Q)$.  Analogously the velocity can be
expressed by \Eqref{Tihomir93Eqn}
\be
\mathbf{v}(\xi)=\mathrm{d}_{Q\xi} \mathbf{b}(\xi),\qquad
\mathrm{d}_{Q\xi}=\frac{\mathrm{d}}{Q\mathrm{d}\xi},
\ee
i.e., plasma's displacement is parameterized by the magnetic field
\be
\mathbf{b}(\xi)=Q\int \mathbf{v} \mathrm{d} \xi
=V_\mathrm{A}k_y \int \mathbf{v} \mathrm{d} t
\ee
recalling that magnetic field lines are frozen into the highly
conductive fluid.

The physical time $t$ is related to the dimensionless one $\tau$
\be
t=\frac{\tau}{A}=\frac{Q\xi}{A K_y}.
\ee
The time-dependence of the wave-vector \Eqref{k(t)} allows us to
express the magnetic field $\mathbf{B}(t,\mathbf{r})$ and velocity
of the fluid $\mathbf{V}(t,\mathbf{r})$ \Eqref{wave}.  The general
solution depends on 4 arbitrary constants $C_\mathrm g,C_\mathrm
u,\tilde C_\mathrm g, \tilde C_\mathrm u.$

After some algebra we can express also the wave component of the
pressure\cite{Mishonov:07}
\be
\label{pressure}
p(t,\mathbf{r})=\rho V_A^2 P(t)\,
\mathrm{e}^{\ii\mathbf{k}(t) \cdot\, \mathbf{r}},\qquad
P(t)=-b_y(t)+2\frac{K_yv_x(t)}{K^2(t)}.
\ee
In such a way we derive a general analytical solution for linearized
MHD waves in a shear flow of a magnetized fluid. In the next
section, we will analyze the behavior of this exact solution when
the waves amplification is significant.

The essential part of our analytical solution for three-dimensional
SMWs in a shear flow is presented by the $x$-component of the
magnetic field
\begin{eqnarray}
&&B_x=B_0\, \mathrm{Re}\left\{\exp\left[\ii\left(\mathbf{k}(t)\cdot\mathbf r
+\varphi_0\right)\right]\right\}\\
&&\qquad\times \left\{C_\mathrm g\,
\mathrm{HeunC}(0,-\frac{1}{2},0,-\frac{Q^2}{4},\frac{1+Q^2}{4},-\xi^2) \right.\nonumber\\
&&\qquad\left.+C_\mathrm u\,
\xi\,
\mathrm{HeunC}(0,+\frac{1}{2},0,-\frac{Q^2}{4},\frac{1+Q^2}{4},-\xi^2)\right\},\nonumber\\
&&\quad=B_0\cos\left[\mathbf{k}(t)\cdot\mathbf r +\varphi_0\right]
\left\{ C_\mathrm g b_{x,\,\mathrm{g}}(\xi)
             +C_\mathrm u b_{x,\,\mathrm{u}}(\xi)\right\}\nn,
\end{eqnarray}
where
\begin{eqnarray}
\xi=\frac{K_y}{Q}\,\tau=\frac{k_y}{\sqrt{k_y^2+k_z^2}} \, A (t-t_0),\\
Q^2=\Lambda^2(k_y^2+k_z^2),\qquad \Lambda^2=\frac{B_0^2}{\mu_0\rho A^2}.
\end{eqnarray}
Here $\varphi_0$ and $t_0$ are arbitrary constants parameterizing
the initial conditions.  All other MHD variables: velocity, pressure
and displacement of the plasma by the wave can be expressed by this
solution. The dimensionless times have simple geometrical
interpretation in the \textbf{k}-space
\be
\tau=-\frac{k_x(t)}{k_y}, \qquad
\xi=-\frac{k_x(t)}{\sqrt{k_y^2+k_z^2}}.
\ee

For the velocity of AWs a substitution of $\psi$ from \Eqref{bxfin}
into \Eqref{bzfin} gives
\be
\label{vzfin}
v_z(\xi)=\mathrm{d}_{Q\xi} b_z= -\tilde{C}_\mathrm g\sin(Q\xi)
+\tilde{C}_\mathrm u\frac{\cos(Q\xi)}{Q}
+\frac{2K_z}{Q}
\int_{-\infty}^{\xi}
\cos[Q(\xi-\xi^{\prime})]\,
\frac{v_x(\xi^{\prime})}
{1+\xi^{\prime 2}}\,\mathrm{d}\xi^{\prime}.
\ee
If we substitute here $\psi$ from Eqs.~(\ref{gen}), (\ref{PsiHeun_g}) and
(\ref{PsiHeun_u}), we arrive at the final analytical solution which
contains Heun functions.

\section{Amplification of the MHD Waves} %
%
The odd and even solutions have asymptotics at $\xi\rightarrow
\infty$
\begin{equation}
\label{asymptotics}
\psi_\mathrm g \approx D_\mathrm g\cos(Q\xi+\delta_\mathrm g),\qquad
\psi_\mathrm u \approx D_\mathrm u\cos(Q\xi+\delta_\mathrm u),
\end{equation}
where the asymptotic phase shifts $\delta_\mathrm g(Q^2),\;\delta_\mathrm u(Q^2)$
depend on the effective energy.
For a sufficiently large wave-vectors we have the asymptotic
\begin{equation}
\delta_\mathrm g(Q^2\gg1)\approx0, \qquad
\delta_\mathrm u(Q^2\gg1)\approx-\frac{\pi}{2}.
\end{equation}

As we will see later, the averaged amplification coefficient of the
energy of MHD waves $\mathcal{A}(\delta_\mathrm g,\delta_\mathrm u)$
depends only on the asymptotic phases of the solutions but not on the
amplitudes $D_\mathrm g$ and $D_\mathrm u$.

\subsection{Auxiliary Quantum Mechanical Problem}%
%
Temporarily introducing imaginary exponents
$\exp(\ii\mathbf{k}\cdot\mathbf{r})$ instead of
$\sin(\mathbf{k}\cdot\mathbf{r})$ and
$\cos(\mathbf{k}\cdot\mathbf{r})$ gives significant simplification
of the analytical calculations related to the physics of the waves. In
this subsection we will consider $\psi$ in \Eqref{Sch} as a complex
function in order to make easier any further analysis of the MHD
amplification coefficient. In order to analyze the effective
MHD equation \Eqref{Sch} we will solve the quantum-mechanical
counterpart of our MHD problem, that is a tunneling through a barrier
$2m U/\hbar^2=1/(1+\xi^2)^2,$ supposing that $\psi$ is a complex
function.
Consider an incident wave with a unit amplitude, a reflected wave with
amplitude $R$ and a transmitted wave with amplitude $T$
\begin{eqnarray}
&&\psi(\xi\rightarrow -\infty)\approx \exp(\ii Q\xi)
  + {R}\exp(-\ii Q\xi),\\
&&\psi(\xi\rightarrow +\infty)\approx  T\exp(+\ii Q\xi).
\end{eqnarray}
Using the asymptotics of the eigen-functions
\be
\label{asg}
\psi_\mathrm g\approx \left\{ \begin{array}{l}
D_\mathrm g\cos(Q\xi-\delta_\mathrm{g})
 \qquad\mbox{for  } \xi\rightarrow -\infty,\\
D_\mathrm g\cos(Q\xi+\delta_\mathrm g)
 \qquad\mbox{for  } \xi\rightarrow +\infty,
\end{array}\right.
\ee
and
\be
\label{asu}
\psi_\mathrm u\approx \left\{ \begin{array}{l}
 -D_\mathrm u\cos(Q\xi-\delta_u)
 \quad\;\;\;\mbox{for  } \xi\rightarrow -\infty,\\
\;\;\, D_\mathrm u\cos(Q\xi+\delta_\mathrm u)
  \qquad\mbox{for  } \xi\rightarrow +\infty
\end{array}\right.
\ee
as well as the general condition
\begin{equation}
\psi(\xi)=C_\mathrm g^\mathrm{(q)} \psi_\mathrm g(\xi)
+ C_\mathrm u^\mathrm{(q)} \psi_\mathrm u(\xi),
\end{equation}
we compare the coefficients in front of $\exp(iQ\xi)$ and $\exp(-iQ\xi)$
for $\xi\rightarrow -\infty$ and $\xi\rightarrow \infty$.  The
solution to a simple matrix problem yields
\be
C_\mathrm g^\mathrm{(q)}D_\mathrm g=\exp(\ii\delta_\mathrm g),\qquad
C_\mathrm u^\mathrm{(q)}D_\mathrm u=-\exp(\ii\delta_\mathrm u)
\ee
Then for the tunneling amplitude we get
\be
T=|T|\mathrm{e}^{\ii(\delta_\mathrm u+\delta_\mathrm g-\pi/2)}
\ee
and finally for the tunneling coefficient we obtain
\be
\mathcal{D}=|T|^2=\mathrm{s_{ug}^2},\qquad
\mathrm{s_{ug}}\equiv\sin(\delta_\mathrm u-\delta_\mathrm g).
\ee
The convenience of the tunneling coefficient is that it varies in
the range $0\le\mathcal{D}\le1$. In the next two subsections we will
present the SMW amplification coefficient as a function of the
tunneling coefficient $\mathcal{K}=2/\mathcal{D}-1$ on the analogy
of Heisenberg's ideas in quantum mechanics where the statistical
properties of the scattering problem depend only on the phases of
the $S$-matrix.

The phases $\delta_\mathrm{g},\;\delta_\mathrm{u}\in (-\pi,\;\pi)$
and amplitudes $D_\mathrm{g}$ and $D_\mathrm{u}$
can be determined continuing the exact wave functions
Eqs.~(\ref{PsiHeun_g})--(\ref{PsiHeun_u}) with WKB asymptotics
\Eqref{asymptotics}. Thus we obtain
\begin{eqnarray}
&&\psi=D\cos(Q\xi+\delta)\,,\qquad
\tilde\delta=-Q\xi-\arctan\frac{\mathrm{d}_\xi\psi(\xi)}{Q\psi(\xi)}\,,
\nn\\
&&\delta=\tilde\delta-2\pi\times
\mathrm{int}\left(\frac{\tilde\delta+\pi}{2\pi}\right)\in (-\pi,\;\pi)\,,
\qquad D=\frac{\psi(\xi)}{\cos(Q\xi+\delta)}\,
,\end{eqnarray}
at some sufficiently large $\xi\gg 1+2\pi/Q$.  Here $\mathrm{int}(\dots)$
stands for the integer part of a real number. When programming we have to
use the two-argument $\arctan$ function
\be
\arctan(y,x)=\arctan(y/x) + \frac{\pi}{2}\,\theta(-x)\,\mathrm{sgn}(y)\in (-\pi,\;\pi).
\ee
The accuracy of this continuation is controlled by the Wronskian
from the asymptotic wave functions
\be
\label{Wronskian}
W(\psi_\mathrm{g},\psi_\mathrm{u})(\xi)
=QD_\mathrm{g}D_\mathrm{u}\sin(\delta_\mathrm{g}-\delta_\mathrm{u})=1.
\ee
%

\subsection{MHD and Real $\psi$}%
%
Imagine that in an ideal plasma we have at $t \to -\infty$ some
plane MHD wave -- our task is to calculate how many times the energy
density increases at $t \to \infty$, and to average this
amplification over all the initial phases of that wave. As
there is no amplification for the $b_z$ component according to
\Eqref{bz} we will concentrate our attention on the $b_x$ component.
The amplification comes from the negative ``friction'' term $\propto
\xi/(1+\xi^2)$ in \Eqref{bx}. The influence of this friction is
transmitted to the effective potential barrier $\propto
1/(1+\xi^2)$.  For $Q^2<1$ we have an analog of the quantum
mechanical tunneling.

In the current MHD problem $\psi$ is a real variable with
asymptotics
\be
\label{wave-asymp}
\psi \approx \left\{ \begin{array}{l}
\quad \; \cos(Q\xi-\phi_\mathrm{i})
 \qquad\mbox{for  } \xi\rightarrow -\infty,\\
D_\mathrm{f}\cos(Q\xi+\phi_\mathrm{f})
 \qquad\mbox{for  } \xi \rightarrow +\infty,
\end{array}\right.
\ee
i.e., we have an incident wave with a unit amplitude and an initial phase
$\phi_\mathrm{i}$. $D_\mathrm{f}$ is the amplitude and
$\phi_\mathrm{f}$ is the phase of the amplified wave.

Again we present the $\psi$ function as linear combination of even
and odd solutions
\begin{equation}
\psi(\xi)=C_\mathrm g^\mathrm{(c)} \psi_\mathrm g(\xi) + C_\mathrm u^\mathrm{(c)}
\psi_\mathrm u(\xi).
\end{equation}
Here we substitute the asymptotic formulas \Eqref{asg} and
\Eqref{asu}, and the comparison of the coefficient with
\Eqref{wave-asymp} at $\xi\rightarrow -\infty$ gives
\be
C_\mathrm{g}^{\mathrm{(c)}}D_\mathrm{g}=
\mathrm{\frac{s_{iu}}{s_{gu}}}
,\qquad
C_\mathrm{u}^{\mathrm{(c)}}D_\mathrm{u}=
\mathrm{\frac{s_{ig}}{s_{gu}}},
\ee
where
\be
\mathrm{s_{iu}}\equiv\sin(\phi_\mathrm{i}-\delta_\mathrm u),\qquad
\mathrm{s_{ig}}\equiv\sin(\phi_\mathrm{i}-\delta_\mathrm g).
\ee
The comparison of the coefficients at $\xi\rightarrow \infty$
gives for the phase and the amplification of the signal
\begin{eqnarray}
&&\phi_\mathrm{f}=F(\phi_\mathrm{i})\equiv
\arctan
\mathrm{\frac{s_{ig} s_{u} + s_{iu} s_{g}}{s_{ig} c_{u} + s_{iu} c_{g}}},\\
&&\mathcal{A}(\phi_\mathrm{i})\equiv D_\mathrm{f}^2=\frac{\mathcal{N}}{\mathcal{D}},
\end{eqnarray}
where
\begin{eqnarray}
&&\mathcal{N}= \mathrm{ (s_{ig} s_{u} + s_{iu} s_{g})^2
 + (s_{ig} c_{u} + s_{iu} c_{g})^2 },\\
&&\mathrm{s_{g}}=\sin(\delta_g),\qquad
\mathrm{s_{u}}=\sin(\delta_\mathrm u),\\
&&\mathrm{c_{g}}=\cos(\delta_g),\qquad
\mathrm{c_{u}}=\cos(\delta_\mathrm u).
\end{eqnarray}
The reversibility of the dissipation-free motion leads us to
$\phi_\mathrm{i}=F(\phi_\mathrm{f})$, i.e., function $F$
coincides with its inverse function $F(F(\phi))=\phi$.  As time
reverses the wave amplification is converted to attenuation
(damping in some sense)
$\mathcal{A}(\phi_\mathrm{i})\mathcal{A}(F(\phi_\mathrm{i}))=1$.

In unabridged mathematical notations we have the function
\be
F(\varphi)\equiv
\arctan \mathrm{
\frac{\sin{(\varphi - \alpha)} \sin{\beta} + \sin{(\varphi - \beta)}\sin{\alpha}}
{\sin{(\varphi - \alpha)} \cos{\beta} + \sin{(\varphi - \beta)} \cos{\alpha}}},
\ee
defined in the interval $\varphi \in (-\pi/2,\; \pi/2).$  For
arbitrary values of the parameters $\alpha$ and $\beta$
\be
F[F(\phi)]=\phi,
\ee
i.e., this function $F$ coincides with its inverse function $F^{-1}$.
The nonlinear function $F$ has only $2$ immovable points
\be
F(\alpha)=\alpha,\qquad F(\beta)=\beta.
\ee
Defining also
\be
\mathcal{A}(\varphi)\equiv\left\{\left[\sin{(\varphi-\alpha)}\sin{\beta}
              +\sin{(\varphi-\beta)}\sin{\alpha}\right]^2
+\left[\sin{(\varphi-\alpha)}\cos{\beta}
              +\sin{(\varphi-\beta)}\cos{\alpha}\right]^2\right\}
              /\sin^2(\alpha-\beta),
\ee
we have another curious relation
\be
\mathcal{A}[F(\varphi)]\,\mathcal{A}(\varphi)=1.
\ee

The so derived amplification coefficient
$\mathcal{A}(\phi_\mathrm{i};\delta_\mathrm g,\delta_\mathrm u)$
depends on the initial phase. In the next subsection we will
consider the statistical problem of phase averaging with respect to
the initial phase $\phi_\mathrm{i}$.

\subsection{Phase Averaged Amplification}%
%
For waves generated by turbulence the initial phase is unknown and
one can suppose a uniform phase distribution. That is why for
solving the statistical problem of energy amplification we need to
calculate average values with respect to the initial phase
$\phi_\mathrm{i}$. That idea is coming from the well-known random
phase approximation (RPA) in plasma physics. The phase averaging
already introduces an element of irreversibility because we already
suppose that waves are created with random phases. This is the MHD
analog of the molecular chaos from the theorem for entropy increase
in the framework of the kinetic theory if the probability distributions
are introduced in the initial conditions of the mechanical problem.
In the case of accretion flows we also suppose that turbulence
is a chaotic phenomenon and we have to apply the RPA for investigating
the statistical properties.

The calculation of the integral
\be
\int_{0}^{\pi}\frac{\mathrm{d}\phi_\mathrm{i}}{\pi}\mathcal{N}(\phi_\mathrm{i})
=2-\mathrm{s_\mathrm{ug}^2}=2-\mathcal{D}
\ee
gives for the initial phase averaged gain
\begin{equation}
\label{ampl-trans}
\mathcal{G} \equiv \int_{-\pi/2}^{\pi/2}
\mathcal{A}(\phi_\mathrm{i}) \, \frac{\mathrm{d}
\phi_i}{\pi} = \frac{2}{\mathrm{
\sin^2(\delta_\mathrm{u}-\delta_\mathrm{g})
}}=\frac{2}{\mathcal D}-1.
\end{equation}
In such a way the SMW amplification coefficient $\mathcal G$ is
presented by the tunneling coefficient $\mathcal D$ of the
corresponding quantum problem. Both coefficients are expressed by
the asymptotic phases $\delta_\mathrm g,$ $\delta_\mathrm u$ in
analogy with partial waves phase analysis of the quantum mechanical
scattering problem in atomic and nuclear physics. The axial symmetry
of this result $\mathcal{G}\left(\sqrt{K_y^2+K_z^2}\right)$
significantly simplifies the further statistical analysis.

Let us analyze the physical meaning of the gain coefficient
$\mathcal{G}$.  As
\begin{eqnarray}
&&Q\xi=K_y\tau=V_\mathrm{A}k_y\,t=\omega_\mathrm{A}\mathrm{sgn}(k_y)t,\qquad
\omega_\mathrm{A}(\mathbf{k})=\left|\mathbf{V}_\mathrm{A}\cdot\mathbf{k}\right|
=V_\mathrm{A}|k_y|\ge 0,\\
&&\mathbf{v}_\mathrm{gr}\equiv\frac{\partial\omega_\mathrm{A}}{\partial \mathbf{k}}
=\mathbf{V}_\mathrm{A}\mathrm{sgn}\left(\mathbf{V}_\mathrm{A}\cdot\mathbf{k}\right)
=V_\mathrm{A}\mathrm{sgn}(k_y)\mathbf{e}_y,
\end{eqnarray}
the asymptotics \Eqref{wave-asymp} let us conclude that for
$|t|\rightarrow\infty$ we have only magnetosonic waves with
dispersion coinciding with that of Alfv\'en waves.\cite{Alfven:42}
In the spirit of M.~T.~Weiss quantum interpretation of the classical
Manley--Rowe theorem\cite{LL8,Zhelyazkov:00} one can present the
wave energy $\hbar \omega_\mathrm{A} N$ by a number of quanta, the
number of \textit{alfvenons}: ``The alfvenons introduced in this
Letter\cite{Stasiewicz:06} appear to be effective and spectacular
converters of electromagnetic energy flux into kinetic energy of
particles.'' We use this notion in a slightly different sense, our
former terminology was \textit{alfvons}\cite{Mishonov:07} in our
case. Following this interpretation, the energy gain $\mathcal{G}$
describes the increasing number of quanta
\be
\mathcal{G}\left(\sqrt{k_y^2+k_z^2}\right)
=\frac{\hbar\omega_\mathrm{A}(|k_y|)
\overline{N}_\mathrm{alfvenons}(t\rightarrow+\infty)}
{\hbar\omega_\mathrm{A}(\left|k_y\right|)
\overline{N}_\mathrm{alfvenons}(t\rightarrow-\infty)},
\ee
as in a laser system. In this terminology the mechanism of heating
of quasars can be phrased, namely due to lasing of alfvenons in
shear flows of magnetized plasma.  Laser or rather
maser\cite{Trakhtengerts:08} effects are typical phenomena in space
plasmas. The hydrodynamic overreflection
instability\cite{Fridman:08} and burst-like increase of the wave
amplitude\cite{Rogava:03} are phenomena of similar kind. More
precisely $\mathcal{G}$ is the gain for the $x$--$y$-polarized SMWs,
the energy of mode conversion in z-polarized AWs will be analyzed
elsewhere. The notion amplification is correct for standing waves
but
$2\cos(\mathbf{k}\cdot\mathrm{r})
=\mathrm{e}^{\mathrm{i}\mathbf{k}\cdot\mathrm{r}}
+\mathrm{e}^{-\mathrm{i}\mathbf{k}\cdot\mathrm{r}}$
and amplification is simultaneous for opposite wave-vectors. That is
why some theorists prefer to use overreflection in spite that there
is no rigid object reflecting the waves.

\subsection{Analytical Approximations for Amplification} %
%

For scattering problems by a localized potential at small
wave-vectors $Q\ll1$, when the wavelength is much larger than the
typical size of the nonzero potential, we can apply the
delta-function approximation
\be
\label{delta-Sch}
\frac{2m}{\hbar^2}U(\xi)\equiv \frac{1}{(1+\xi^2)^2}\rightarrow
2Q_0\delta(\xi).
\ee
In this well-known quantum mechanical problem\cite{Greiner:01} the
transmission coefficient is
\be \label{D-E} D\approx\frac{Q^2}{Q^2+Q_0^2},\qquad
\delta_\mathrm{u}=-\frac{\pi}{2},\qquad
\delta_\mathrm{g}=\frac{\pi}{2}+\arctan\frac{Q}{Q_0},
\qquad \psi_\mathrm{g}
=\frac{\cos(|Q\xi|+\delta_\mathrm{g})}{\cos(\delta_\mathrm{g})},
\qquad \psi_\mathrm{u}=\frac{\sin Q\xi}{Q}
.\ee

According to the tradition of the method of potential of zero
radius, the coefficient $Q_0\approx \frac{1}{2}\pi$ is determined by
the behavior of the phases at small wave-vectors. Only qualitatively
this parameter corresponds to the area of the potential
\be
2Q_0=\frac{\pi}{2}Z_\mathrm{ren}=\pi,\qquad
\frac{\pi}{2}=\int_{-\infty}^{\infty} \frac{1}{(1+\xi^2)^2}
\,\mathrm{d} \xi,
\ee
but the renormalizing coefficient $Z_\mathrm{ren} = 2$ which we have
to introduce differs from unity. According to the general relation
\Eqref{ampl-trans} for the amplification we have
\be \label{EnergyGain}
\mathcal{G}-1=2\left(\frac{1}{\mathcal{D}}-1\right)
\approx\frac{2Q_0^2}{Q^2}, \qquad 2Q_0^2= \frac{\pi^2}{2}\approx
4.934\approx 5\,. \ee

\begin{figure}
\centering\includegraphics{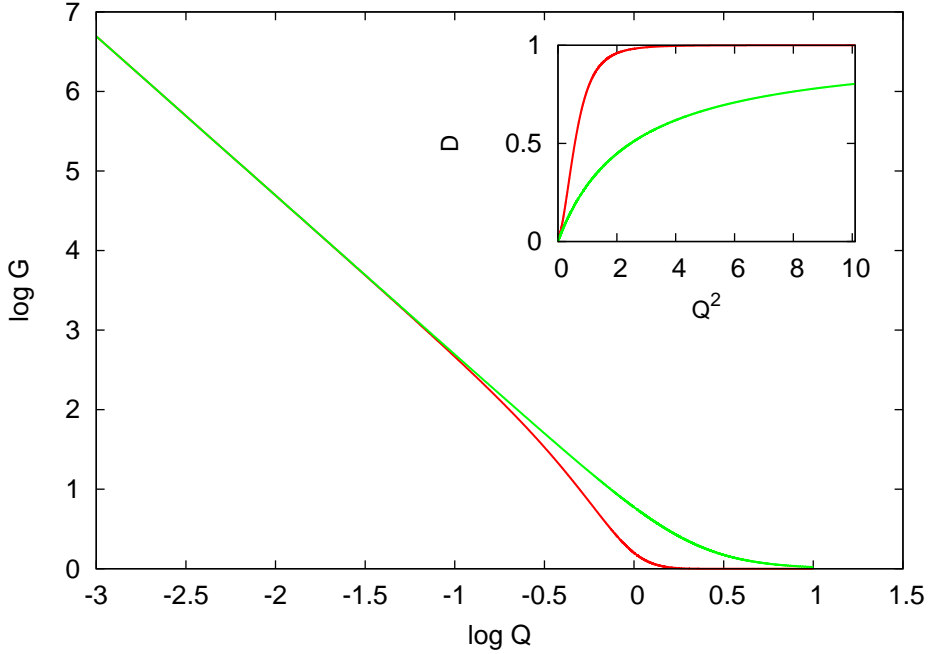} \caption{The efficiency of the dissipation
mechanism that is likely responsible for the formation of stars and other
compact astrophysical
objects: logarithm of the wave energy amplification
$\mathcal{G}=2/\mathcal{D}-1$ (cf.\ with \Eqref{ampl-trans})
versus the logarithm of the dimensionless
wave-vector $Q$. The delta-function approximation \Eqref{EnergyGain}
(\textcolor{green}{upper curve}) is asymptotically exact at huge (giant)
amplifications. In the inset the transmission coefficient
$\mathcal{D}$ is plotted versus the square of the wave-vector $Q^2$ of the
auxiliary quantum mechanical problem. The delta-function potential
approximation (\textcolor{green}{lower curve}) \Eqref{D-E} is good only
for long enough wavelengths $Q^2\ll1$ at which the amplification is
significant $\mathcal{G}\gg1$; $\log \mathcal G= 6$ means energy
amplification one million times or $60$~dB.
\label{fig:gain}
}
\end{figure}

The scattering phases $\delta_\mathrm g$ and $\delta_\mathrm u$ can
be obtained by a fit of the asymptotic wave functions \Eqref{asg}
and \Eqref{asu} with the analytical solutions \Eqref{PsiHeun_g} and
\Eqref{PsiHeun_u}. In such a way we can calculate the wave-vector
dependence of the transmission coefficient as it is depicted in
\Fref{fig:gain}.

Let us derive the analytical formula for $Q_0$: The solutions
\begin{eqnarray}
&\psi_g(\xi)& =\sqrt{1+\xi^2} , \\
&\psi_u(\xi)& = \sqrt{1+\xi^2} \arctan(\xi)
\end{eqnarray}
of effective Schr\"odinger's equation \Eqref{Sch} in long wavelength
limit $Q\rightarrow0$
\be \label{Sch0} \mathrm{d}_\xi^2\psi-\frac{1}{(1+\xi^2)^2}\psi=0
\ee
with asymptotics at $\xi\rightarrow\infty$
\begin{eqnarray}
\psi_g(\xi\gg 1) \approx \xi, \\
\psi_u(\xi\gg 1) \approx \frac{\pi}{2}\xi
\end{eqnarray}
have to  be compared with the approximative solutions \Eqref{asg}
and \Eqref{asu} for $\xi\gg1,$ when for $Q\rightarrow0$ we have
$|\delta_g|\approx|\delta_u|\approx\frac{\pi}{2}$ and sinusoids are
almost linear
\begin{eqnarray}
\psi_g(1 \ll \xi \ll \frac{1}{Q}) \approx D_g\sin(Q\xi) \approx D_g Q\xi, \\
\psi_u(1 \ll \xi \ll \frac{1}{Q}) \approx D_u\sin(Q\xi) \approx D_u
Q\xi.
\end{eqnarray}
The comparison of the first derivatives in this region
\begin{eqnarray}
\mathrm{d}_\xi \psi_g(\xi\gg 1) \approx 1 \approx D_g Q,\\
\mathrm{d}_\xi \psi_u(\xi\gg 1) \approx \frac{\pi}{2}\approx D_u Q
\end{eqnarray}
gives
\be D_g\approx\frac{1}{Q}, \qquad D_u\approx \frac{\pi}{2Q}. \ee
Then formula for Wronskian \Eqref{Wronskian} determines the phase
difference
\be \sin(\delta_\mathrm{g}-\delta_\mathrm{u})\approx\frac{2}{\pi}Q
\ee
and transmission coefficient
\be
\mathcal{D}\approx\sin^2(\delta_\mathrm{g}-\delta_\mathrm{u})=\left(\frac{2Q}{\pi}\right)^2.
\ee
The \Eqref{ampl-trans} then gives for the phase averaged
amplification \be
G=\frac{2}{D}-1\approx\frac{\pi^2}{2Q^2}+\mathrm{const} \ee and the
comparison with $\delta$-potential approach \Eqref{EnergyGain} gives
the analytical result for the strength of the $\delta$-potential
$Q_0=\pi/2$ given in \Eqref{EnergyGain}.

Our problem formally coincides with the quantum problem\cite{Landau:32} of
transmission coefficient $\mathcal{D}\propto E$ at low energies
\be
U(x)=\frac{U_0}{[1+(\alpha x)^2]^2},\quad
U_0=\frac{\hbar^2\alpha^2}{2m},\quad
E= U_0 Q^2, \quad
D\approx \left(\frac{2Q}{\pi}\right)^2=\frac{4}{\pi^2}\frac{E}{U_0}\ll 1.
\ee

The zero-radius potential Pad\'e approximant \Eqref{D-E} has an
acceptable accuracy for small wave-vectors. Having the quantum
mechanical transmission coefficient, we can calculate the energy
gain coefficient $\mathcal{G}-1$. In the next section we will
incorporate the approximative solution
\be
\label{gain}
\mathcal{G}-1\sim\frac{1}{Q^2}
=\frac{A^2}{V^2_\mathrm{A}q^2},\qquad
q^2= k_\perp^2\equiv k_y^2+k_z^2=\mathrm{const}
\ee
in the simplest model of Kolmogorov turbulence. Here the
subscript $\perp$ means that the amplification depends on the projection of the
wave-vector perpendicular to the shear velocity and magnetic field.

The delta function approximation has a visual interpretation in
classical mechanics as well, supposing that $\psi$ is the
displacement of an oscillator and $\xi$ is the time. The
approximative equation
\be
\mathrm{d}_\xi^2\psi=-Q^2\psi(\xi) + 2Q_0\delta(\xi)\psi(\xi)
\ee
means that at a time moment $\xi=0$ the oscillator is
subjected to a forcing impulse with a magnitude
\be
\label{jump}
\mathrm{d}_\xi\psi(+0)- \mathrm{d}_\xi\psi(-0)=2Q_0\psi(0),
\qquad\mbox{or}\qquad
v_x(+0)- v_x(-0)=\frac{\pi\,b_x(0)}{\sqrt{K_y^2+K_z^2}},\qquad
\mbox{when}\quad K_x(0)=0.
\ee
If for $Q\ll Q_0$ we have initial oscillations with amplitude $A_\mathrm{i}$
\be \psi=A_\mathrm{i}\cos Q\xi\qquad\mbox{for}\quad \xi<0, \ee
after the push in $\xi=0$ we have oscillations with much increased
amplitude
\be
\psi=A_\mathrm{f}\sin Q\xi,\qquad\mbox{for}\quad \xi>0,\qquad
A_\mathrm{f}\approx \frac{2Q_0}{Q} A_\mathrm{i}\gg A_\mathrm{i}.
\ee
The strong push with an appropriate phase ``amplifies'' the
oscillations; this burst-like increase of the wave amplitude was
observed in numerical investigations of linearized two-dimensional
MHD equations.\cite{Chagelishvili:93,Rogava:03} This phenomenon is
akin to the extremely strong hydrodynamic instabilities due to a
velocity jump; its prediction and discovery both in theory and
experiments are described in Ref.~[\onlinecite{Fridman:08a}]. In
such a way the Alfv\'en's idea of the importance of MHD waves in the
transfer of momentum is reduced to the very simple mathematics for
the jump of the velocity of SMWs \Eqref{jump}. It is instructive to
rewrite this force in the $\mathbf{r}$-space.

\section{Incorporation of Turbulence as Random Driver of MHD Waves}%
%
Being as efficient as has been demonstrated above, the amplification of SMWs,
\Eqref{gain},
is possibly the dominant physical factor responsible for generating and
maintaining the turbulence in accretion flows.  However the theory of turbulence
is much more
complicated than the theory of linearized waves. That is why here we
provide an illustration how the wave amplification can be
incorporated in the turbulence theory. In order to establish common
set of notions and notations we will recall some basic properties of
the homogeneous isotropic Kolmogorov--Obukhov turbulence.

\subsection{Kolmogorov Turbulence}%
%
Let the velocity be presented by the Fourier integral
\be
\mathbf{V}(\mathbf{r})=\int \mathrm{e}^{\ii\mathbf{k}\cdot\mathbf{r}}
\mathbf{V}_\mathbf{k} \frac{\mathrm{d}^3k}{(2\pi)^3},\qquad
\mathbf{V}_\mathbf{k}
=\int\mathbf{V}(\mathbf{r}) \mathrm{e}^{-\ii\mathbf{k}\cdot\mathbf{r}}
\mathrm{d}^3x.
\ee
The energy per unit mass is
\be
\int \frac{1}{2} V^2(\mathbf{r})\mathrm{d}^3x
=\int \frac{1}{2} V^2_\mathbf{k} \frac{\mathrm{d}^3k}{(2\pi)^3}
=\int \mathcal{E}_\mathbf{k}\frac{\mathrm{d}^3k}{(2\pi)^3},
\ee
where we introduce the spectral density averaged with respect to the
turbulence
\be \mathcal{E}_\mathbf{k}
=\frac{1}{2}\langle V^2_\mathbf{k}\rangle_\mathrm{turbulence}.
\ee
We can introduce also the energy of vortices which contains
Fourier components with wavelength $2\pi/k,$ shorter than some fixed
length $\lambda$
\be
\frac{1}{2}V_\lambda^2
=\int_{k<1/\lambda} \mathcal{E}_k \frac{\mathrm{d}^3k}{(2\pi)^3},
\ee
where for isotropic turbulence $\mathcal{E}_\mathrm{k}
=\mathcal{E}_k$. This energy evaluates turbulent pulsation with size
$\lambda$. Further on we will continue with only order of magnitude
evaluations, hence in the following estimations we will drop
off factors such as $4\pi,$ $\frac{1}{2}$, etc.

According to the Kolmogorov--Obukhov (KO) scenario in the inertial
range the magnitude of the velocity pulsations $V_\lambda$ can
depend only on the turbulent power dissipated per unit mass
$\varepsilon$. There is only one combination with the appropriate
dimension
\be
\label{Kolmogorov}
\varepsilon=
\frac{V_\lambda^2}{\lambda/V_\lambda}
=\mathrm{\frac{energy/mass}{time=length/velocity}=\frac{power}{mass}},
\ee
which yields
\begin{eqnarray}
&&V_\lambda^2\sim\left(\varepsilon\lambda\right)^{2/3}
\sim\int_{k\lambda>1}\mathcal{E}_\mathbf{k}^\mathrm{KO}\mathrm{d}^3k
\sim\int_{k\lambda>1}\frac{\varepsilon^{2/3}}{k^{5/3}}\mathrm{d}k,\qquad
\mathrm{d}^3k\sim k^2\mathrm{d}k,\nn \\
&&E(k)=\int k^2 \mathcal{E}_\mathbf{k}^\mathrm{KO} \mathrm{d} \Omega \!
\approx C_\mathrm{K}\varepsilon^{2/3}k^{-5/3},\qquad C_\mathrm{K}\approx 1.6,
\qquad
\mathcal{E}_\mathbf{k}^\mathrm{KO}\sim\frac{\varepsilon^{2/3}}{k^{11/3}}\,.
\end{eqnarray}
$V_\lambda$ is the amplitude of variation in the velocity pulsation
at distance $\lambda$.  $\mathcal{E}_\mathbf{k}$ is the energy
density in the \textbf{k}-space per unit mass; in the
Kolmogorov--Obukhov picture this is a static variable.

The scaling law $ V_\lambda\sim(\varepsilon \lambda)^{1/3}$
\Eqref{Kolmogorov} is applicable for large enough distances
$\lambda>\lambda_0$, where $\lambda_0$ describes the scale where
dissipation effects become essential
\begin{eqnarray}
&&\lambda_0\equiv(\nu_\mathrm k^3/\varepsilon)^{1/4},\qquad
\nu_\mathrm{k}= \eta/\rho, \qquad \rho= M N_p,\qquad N_p=N_e,\qquad
\eta=0.4\frac{M^{1/2}T_p^{5/2}}{e^4\mathcal{L}_p},\qquad\\
&&\mathcal{L}_p=\ln\left(\frac{\lambda_{\mathrm D}T_p}{e^2}\right),\qquad
\frac{1}{\lambda_\mathrm D^2}=4\pi e^2\left(\frac{N_e}{T_e}+\frac{N_p}{T_p}\right),
\qquad e^2=\frac{q_e^2}{4\pi\varepsilon_0},\qquad
\mathcal{L}_e=\ln\left(\frac{\lambda_{\mathrm D}T_e}{e^2}\right),\qquad\nn\\
&&p = N_pT_p + N_eT_e,\qquad
\frac{M}{m}\approx1836,\qquad
c_{\rm s}^2= \frac{5}{3}\frac{T_p+T_e}{M}=\frac{5p}{3\rho},\qquad
\nu_\mathrm{k}=\frac{\eta}{\rho}=\frac{0.4\,T_p^{5/2}}{e^4N_pM^{1/2}\mathcal{L}_p},
\qquad\nn\\&&
\kappa=\frac{0.9\,T_e^{5/2}}{e^4m^{1/2}\mathcal{L}_e},\qquad
C_V=\frac{3}{2}\left(N_e+N_p\right),\qquad
\chi=\frac{\kappa}{C_V},\qquad
\mathrm{P}=\frac{\nu_\mathrm{k}}{\chi}=1.3\frac{\mathcal{L}_e}
{\mathcal{L}_p}\left(\frac{m}{M}\right)^{1/2}
\left(\frac{T_p}{T_e}\right)^{5/2}\ll1,
\nn\\
&&\nu_\mathrm m=\varepsilon_0 c^2 \tilde{\varrho}
=\frac{e^2c^2m^{1/2}\mathcal{L}_e}{0.6\times 4\pi T_e^{3/2}},\qquad
\tilde\varrho=\frac{1}{4\pi\varepsilon_0}
\frac{e^2m^{1/2}\mathcal{L}_e}{0.6T_e^{3/2}},\qquad
\mathrm{P_m}=\frac{\nu_\mathrm{k}}{\nu_\mathrm{m}}
=\frac{0.24\times 4\pi T_e^{3/2}T_p^{5/2}}{mc^2 \, e^6 N_p}
\gg 1.\nn
\end{eqnarray}
Here, apropos, we introduced self-explanatory notations for the
parameters of fully ionized hydrogen plasma:
bare plasma viscosity $\eta$, the
Coulomb logarithm for protons $\mathcal{L}_p$,
the proton mass $M$,
mass density $\rho=MN_p$,
pressure $p$,
sound speed $c_\mathrm{s}$,
magnetic diffusivity $\nu_\mathrm m$,
kinematic viscosity $\nu_\mathrm k$,
heat conductivity $\kappa,$
specific heat per unit volume $C_v,$
temperature conductivity $\chi,$
Prandtl number $\mathrm P,$
magnetic Prandtl number $\mathrm{P_m},$
total viscosity determining Alfv\'en waves damping $\nu=\nu_\mathrm k+\nu_\mathrm m,$
Ohmic resistivity $\tilde\varrho,$
constant in Coulomb interaction $e^2,$
Debye screening length $\lambda_\mathrm D,$
electron temperature $T_e$ and proton temperature $T_p$ times
the Boltzmann constant,
number of electrons and protons per unit volume
$N_e=N_p$, etc.
Those formulas are system invariant: in SI
$\mu_0=4\pi\times10^{-7}=1/c^2\varepsilon_0$, in Gaussian system
$\mu_0=4\pi=1/\varepsilon_0$, while in the Heaviside--Lorentz system
$\mu_0=1=1/\varepsilon_0$.

Let us now consider a magnetosonic wave with a time-dependent
wave-vector \Eqref{k(t)}
\be
\label{q(t)}
q_x(t)=-A(t-t_0)q_y,\qquad q_y=\mathrm{const}, \qquad q_y=\mathrm{const},
\ee
and time-dependent energy density per unit mass in real space
\be
\label{w}
w(t)=\frac{1}{2}
\langle \mathbf{V}_\mathrm{wave}^2+\frac{\mathbf{B}_\mathrm{wave}^2}{\rho\mu_0} \rangle
=\frac{V_\mathrm{A}^2}{4}
\left[\mathbf{b}^2+\left(\mathrm d_{Q\xi} \mathbf{b}\right)^2\right],
\ee
where $\langle\dots\rangle$ stands for spatial averaging,
$\langle\cos^2(\mathbf{k}\cdot\mathrm{r}) \rangle=\frac{1}{2}$. Then
the energy density in the \textbf{k}-space is
\be
\label{density_of_rain}
\mathcal{E}_\mathbf{k}(t)=w(t)\delta[\mathbf{k}-\mathbf{q}(t)].
\ee
Let us mention that all MHD variables $\mathbf{b}$, $\mathbf{v}$,
$w$ in \Eqref{w}, and $P$ in \Eqref{pressure} depend on the
effective wave functions $\psi$ and $\chi$ (solutions to the
effective Schr\"{o}dinger equations) through the dimensionless time
$\xi$:
\begin{eqnarray}
\label{long_system}
&&b_x=\frac{\psi(\xi)}{\sqrt{1+\xi^2}},\\
&&b_y= -\frac{2K_z^2}{K_yQ} \int_{-\infty}^{\xi}
\sin[Q(\xi-\xi^{\prime})]
\frac{v_x(\xi^{\prime})}
{1+\xi^{\prime 2}}\mathrm{d}\xi^{\prime}
+\frac{Q}{K_y}\frac{\xi\psi(\xi)}{\sqrt{1+\xi^2}}
-\frac{K_z}{K_y}\chi(\xi),\\
&&b_z
= \frac{2K_z}{Q} \int_{-\infty}^{\xi}
\sin[Q(\xi-\xi^{\prime})]
\frac{v_x(\xi^{\prime})}
{1+\xi^{\prime 2}}\mathrm{d}\xi^{\prime}+\chi(\xi),
\\
&&v_x=\frac{(1+\xi^2)\mathrm{d}_\xi \psi(\xi) - \xi\psi(\xi)}{Q(1+\xi^2)^{3/2}},\\
&&v_y=-\frac{2K_z^2}{K_yQ} \int_{-\infty}^{\xi}
\cos[Q(\xi-\xi^{\prime})]
\frac{v_x(\xi^{\prime})}
{1+\xi^{\prime 2}}\mathrm{d}\xi^{\prime}
+\frac{\xi(1+\xi^2)d_\xi \psi(\xi) - \xi^2\psi(\xi)}{K_y(1+\xi^2)^{3/2}}
-\frac{K_z}{Q K_y}\,\mathrm{d}_\xi\chi(\xi),
\\
&&v_z
=\frac{2K_z}{Q} \int_{-\infty}^{\xi}
\cos[Q(\xi-\xi^{\prime})]
\frac{v_x(\xi^{\prime})}
{1+\xi^{\prime 2}}\mathrm{d}\xi^{\prime}
+\frac{\mathrm{d}_\xi\chi}{Q}.
\label{long_system_last}
\end{eqnarray}
For programming one can use also formulae explicitly expressed by
the initial conditions $\mathbf{b}(t_0),$ $\mathbf{v}(t_0)$
\begin{eqnarray}
\label{long_system_simpl}
&&b_x=\frac{\psi(\xi)}{\sqrt{1+\xi^2}},
\qquad \xi=-\frac{K_x(\tau)}{\sqrt{K_y^2+K_x^2}}
=-\frac{K_x(\tau_0)}{Q}
+\frac{K_y}{Q}(\tau-\tau_0),
\qquad K_x(\tau)=K_x(\tau_0)-K_y(\tau-\tau_0)
,\\
&&b_y= -\frac{2K_z^2}{QK_y}
\left[\sin(Q\xi)I_\mathrm{c}(\xi)-\cos(Q\xi)I_\mathrm{s}(\xi)\right]
+\frac{Q}{K_y}\frac{\xi\psi(\xi)}{\sqrt{1+\xi^2}}
-\frac{K_z}{K_y}\chi(\xi)
=\frac{Q}{K_y}\,\xi b_x-\frac{K_z}{K_y}\,b_z
,\\
&&b_z = \frac{2K_z}{Q}
\left[\sin(Q\xi)I_\mathrm{c}(\xi)-\cos(Q\xi)I_\mathrm{s}(\xi)\right]
+\chi(\xi),
\\
&&v_x=\frac{(1+\xi^2)\mathrm{d}_\xi \psi(\xi)
- \xi\psi(\xi)}{Q(1+\xi^2)^{3/2}}
=\mathrm{d}_{Q\xi} b_x,\\
&&v_y=-\frac{2K_z^2}{QK_y}
\left[\cos(Q\xi)I_\mathrm{c}(\xi)+\sin(Q\xi)I_\mathrm{s}(\xi)\right]
+\frac{\xi(1+\xi^2)d_\xi \psi(\xi) -
\xi^2\psi(\xi)}{K_y(1+\xi^2)^{3/2}}
-\frac{K_z}{K_y}\,\mathrm{d}_{Q\xi}\chi(\xi)\nn\\
&&\quad =\frac{Q}{K_y}\,\xi v_x -\frac{K_z}{K_y}\,v_z,
\\
&&v_z =\frac{2K_z}{Q}
\left[\cos(Q\xi)I_\mathrm{c}(\xi)+\sin(Q\xi)I_\mathrm{s}(\xi)\right]
+\mathrm{d}_{Q\xi}\chi,
\end{eqnarray}
where
\begin{eqnarray}
&& I_\mathrm{c}(\xi)\equiv\int_{\xi_0}^{\xi}
\cos(Q\xi^{\prime})\,
\frac{v_x(\xi^{\prime})}
{1+\xi^{\prime 2}}\,\mathrm{d}\xi^{\prime}
=\int_{\xi_0}^{\xi}
\cos(Q\xi^{\prime})\,
\frac{(1+\xi^{\prime 2})\mathrm{d}_{\xi^{\prime}} \psi(\xi^{\prime})
- \xi^{\prime}\psi(\xi^{\prime})}{Q(1+\xi^{\prime 2})^{5/2}}
\mathrm{d}\xi^{\prime},\\
&&I_\mathrm{s}(\xi)\equiv\int_{\xi_0}^{\xi}
\sin(Q\xi^{\prime})\,
\frac{v_x(\xi^{\prime})}
{1+\xi^{\prime 2}}\,\mathrm{d}\xi^{\prime}
=\int_{\xi_0}^{\xi}
\sin(Q\xi^{\prime})\,
\frac{(1+\xi^{\prime 2})\mathrm{d}_{\xi^{\prime}} \psi(\xi^{\prime})
- \xi^{\prime}\psi(\xi^{\prime})}{Q(1+\xi^{\prime 2})^{5/2}}
\mathrm{d}\xi^{\prime},\\
&&\chi(\xi)=b_z(\xi_0)\cos[Q(\xi-\xi_0)]+v_z\sin[Q(\xi-\xi_0)],
\qquad \xi_0=-\frac{k_x(t_0)}{\sqrt{k_y^2+k_z^2}}
=-\frac{K_x(\xi_0)}{Q},\\
&& \mathrm{d}_{Q\xi}\chi(\xi)=
-b_z(\xi_0)\sin[Q(\xi-\xi_0)]+v_z\cos[Q(\xi-\xi_0)],
\qquad K_x(\xi)=-Q\xi=K_x(\xi_0)+Q(\xi-\xi_0),
\\
&&\psi(\xi_0)=\sqrt{1+\xi_0^2}\,b_x(\xi_0),\qquad 
\mathrm{d}_\xi\psi(\xi_0)=Q\sqrt{1+\xi_0^2}\,v_x(\xi_0)+
\frac{\xi_0}{\sqrt{1+\xi_0^2}}\,b_x(\xi_0),\\
&&
\psi(\xi)=\left[\psi(\xi_0)\mathrm{d}_\xi\psi_\mathrm{u}(\xi_0)
-\psi_\mathrm{u}(\xi_0)\mathrm{d}_\xi\psi(\xi_0)\right]\psi_\mathrm{g}(\xi)
+\left[\psi_\mathrm{g}(\xi_0)\mathrm{d}_\xi\psi(\xi_0)
-\psi(\xi_0)\mathrm{d}_\xi\psi_\mathrm{g}(\xi_0)\right]\psi_\mathrm{u}(\xi)
.\label{long_system_last_simpl}
\end{eqnarray}
In longwavelength limit $Q\ll1$ and $\xi_0\rightarrow -\infty$ the
numerical integration gives
\begin{eqnarray}
&&J_\mathrm{s,g}
\equiv\int_{-\infty}^{\infty}
\sin(Q\xi^{\prime})\,
\frac{(1+\xi^{\prime 2})\mathrm{d}_{\xi^{\prime}} \psi_\mathrm{g}(\xi^{\prime})
- \xi^{\prime}\psi_\mathrm{g}(\xi^{\prime})}{Q(1+\xi^{\prime 2})^{5/2}}
\mathrm{d}\xi^{\prime}
\approx-0.302\sim 1\,, \\
&&J_\mathrm{c,u} \equiv\int_{-\infty}^{\infty} \cos(Q\xi^{\prime})\,
\frac{(1+\xi^{\prime 2})\mathrm{d}_{\xi^{\prime}}
\psi_\mathrm{u}(\xi^{\prime}) -
\xi^{\prime}\psi_\mathrm{u}(\xi^{\prime})}{Q(1+\xi^{\prime
2})^{5/2}} \mathrm{d}\xi^{\prime} \approx \frac{C_\mathrm{c,u}}{Q}\,
\gg |J_\mathrm{s,g}|,
\qquad C_\mathrm{c,u}\approx 1.459\sim 1\, .
\end{eqnarray}
Due to odd integrants $J_\mathrm{c,g}=0=J_\mathrm{s,u};$
$\lim_{Q\rightarrow 0}[QD_\mathrm{u}]=1.571.$

The formula for $b_z$ describes mutual transformation between MHD
waves with orthogonal polarization. Only for $K_z=0$ we have exact
separation between SMWs with $xy$-polarization and AWs with
$z$-polarization.

An illustration for the initial conditions $\psi(-100)=1,$
$\mathrm{d}_\xi\psi(-100)=0,$
$\chi(-\infty)=0=\mathrm{d}_\xi\chi(-\infty)$ and parameters
$K_y=0.3$, $K_z=0.1$, is presented at the figures below. For
reliability and check of formulae the figures are doubled by the
numerical solution of the set \Eqref{Tihomir93Eqn} by Runge--Kutta
method.\cite{Hairer93}
\begin{figure}
\centering\includegraphics{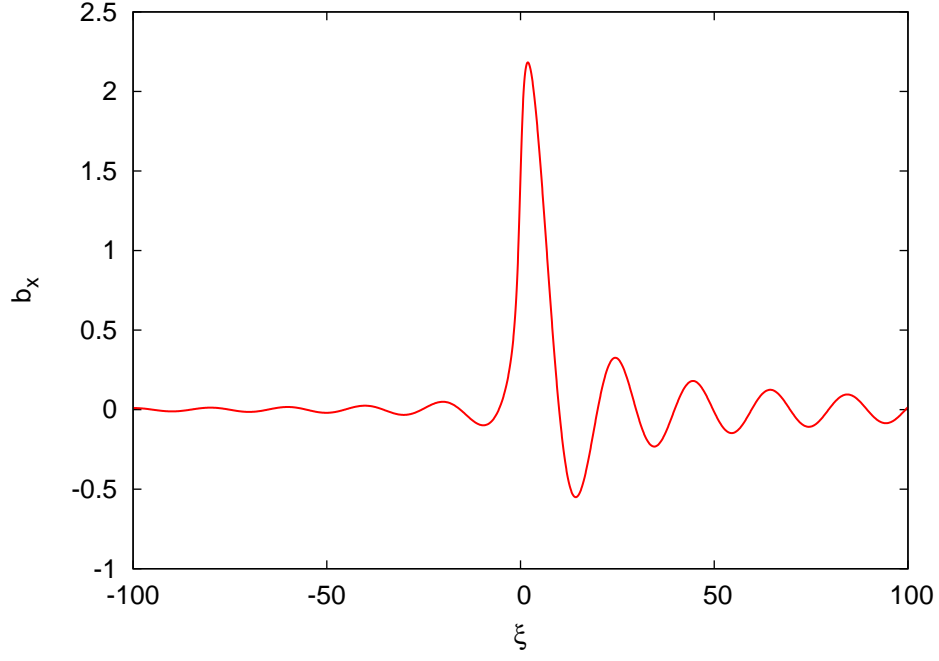} \caption{Radial wave component of
the magnetic field $b_x(\xi)$ as function of the dimensionless time
$\xi$.
\label{fig:bx} }
\end{figure}
\begin{figure}
\centering\includegraphics{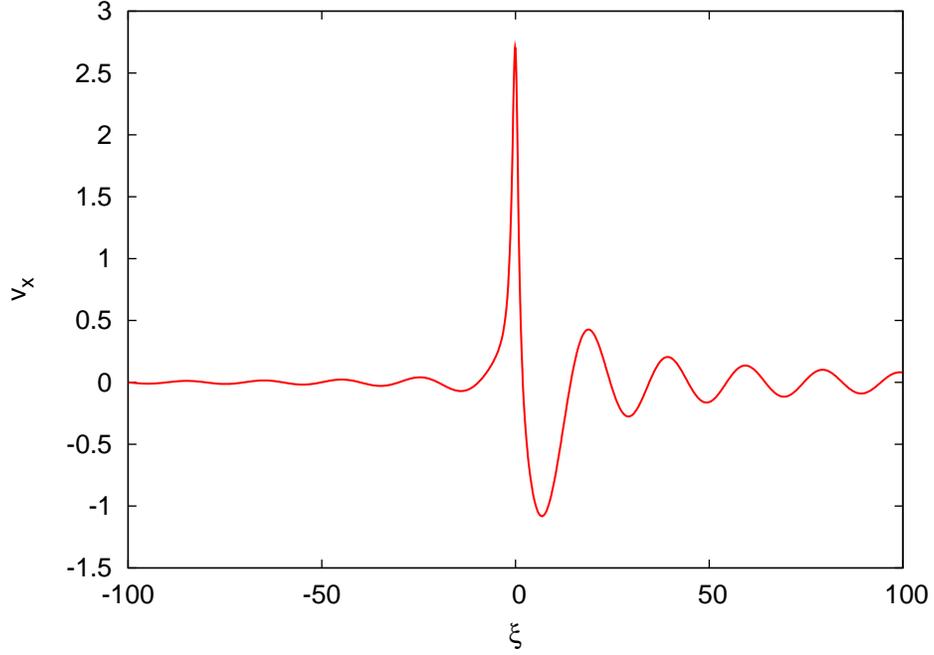} \caption{Time dependenve of the
radial velocity $v_x(\xi)$.
\label{fig:vx} }
\end{figure}
\begin{figure}
\centering\includegraphics{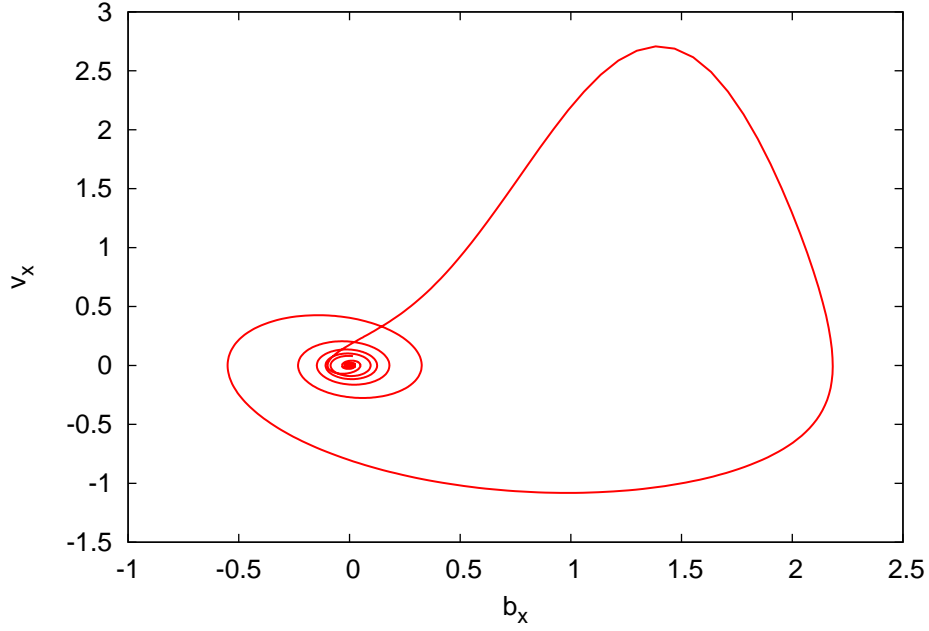} \caption{Phase portrait of
the radial $\hat r=\mathbf{e}_x$ motion $v_x$ versus $b_x.$ The big
loop outlined at $\xi=0$ is the driving force of SMW amplification.
This kick amplifies magnetosonic waves, but the radial amplitude
disappears at $\xi=100.$ The curve starts and ends at the point
($0,0$).
\label{fig:bx_vs_vx} }
\end{figure}
\begin{figure}
\centering\includegraphics{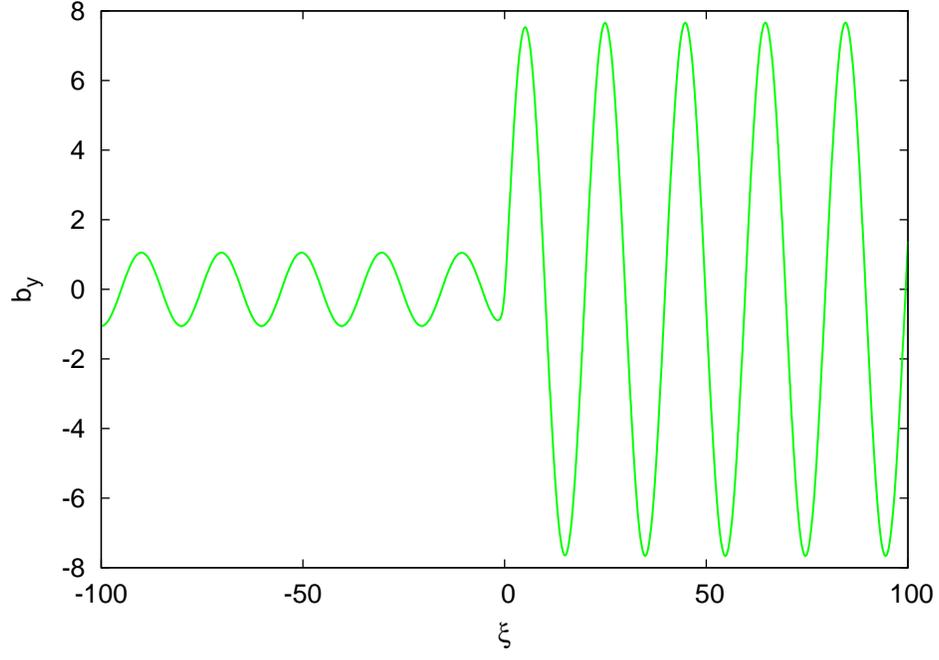} \caption{Wave component of the
magnetic field parallel to the constant one
$\mathrm{B}_0=B_0\mathrm{e}_y$ as function of the dimensionless time
$b_y(\xi)$.
\label{fig:by} }
\end{figure}
\begin{figure}
\centering\includegraphics{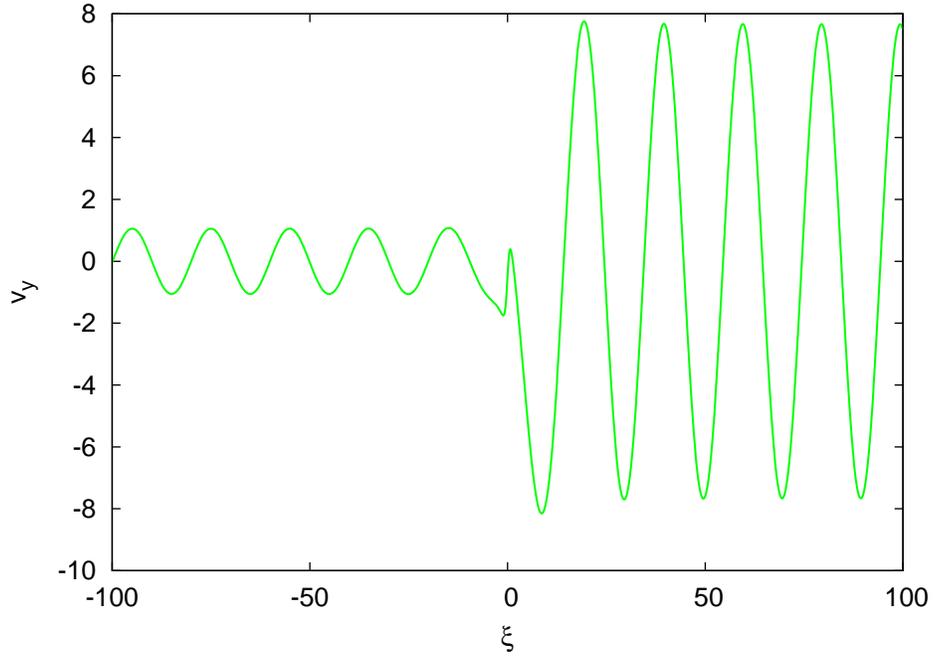} \caption{Longitudinal with
respect to the magnetic field $\mathrm{B}_0$ velocity oscillations
as function of the time $v_y(\xi)$.
\label{fig:vy} }
\end{figure}
\begin{figure}
\centering\includegraphics{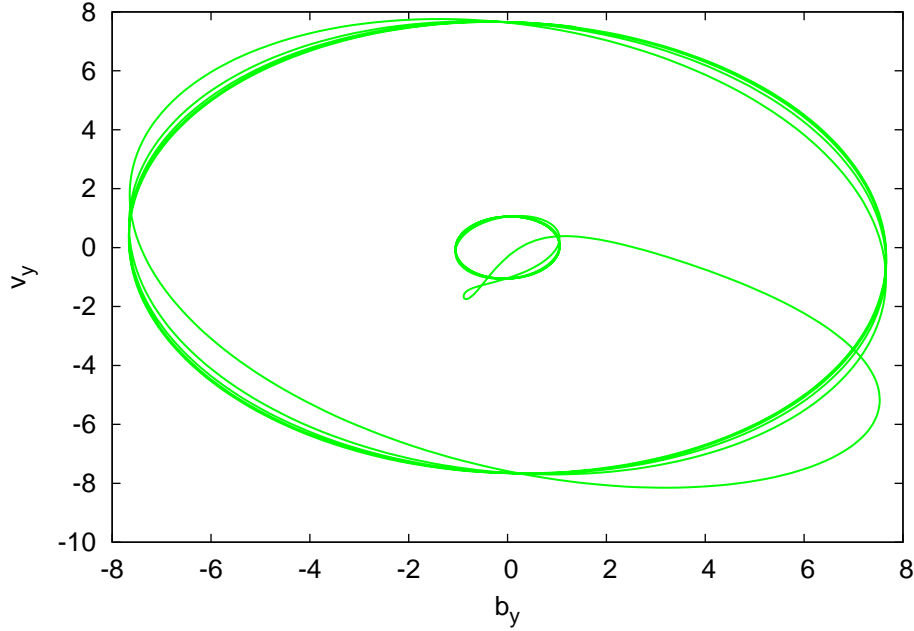} \caption{Phase plot $v_y$
versus $b_y$ of axial $\hat \varphi= \mathbf{e}_y$ SMW oscillations
parallel to the constant magnetic field $B_0\hat \varphi.$ The wave
amplification is the ratio of the  cycle areas at $\xi \to \infty$ and
$\xi \to -\infty.$ The fast transition between those two orbits
describes the lazing of the alfv\'enons.
\label{fig:vy_vs_by} }
\end{figure}
\begin{figure}
\centering\includegraphics{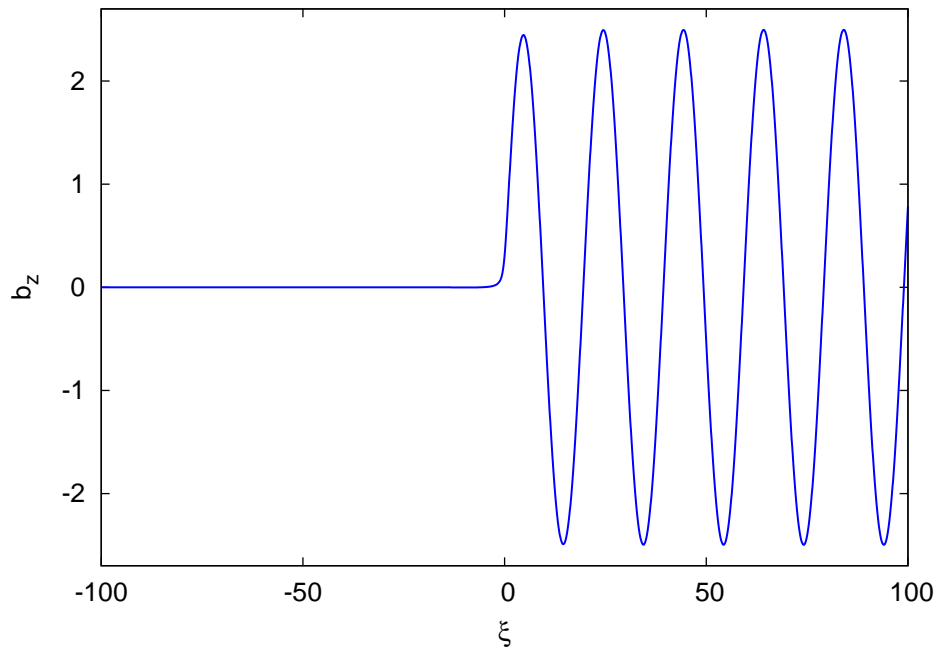} \caption{Time dependence of the
magnetic field oscillations $b_z(\xi)$ perpendicular to the disk
plane.
\label{fig:bz} }
\end{figure}
\begin{figure}
\centering\includegraphics{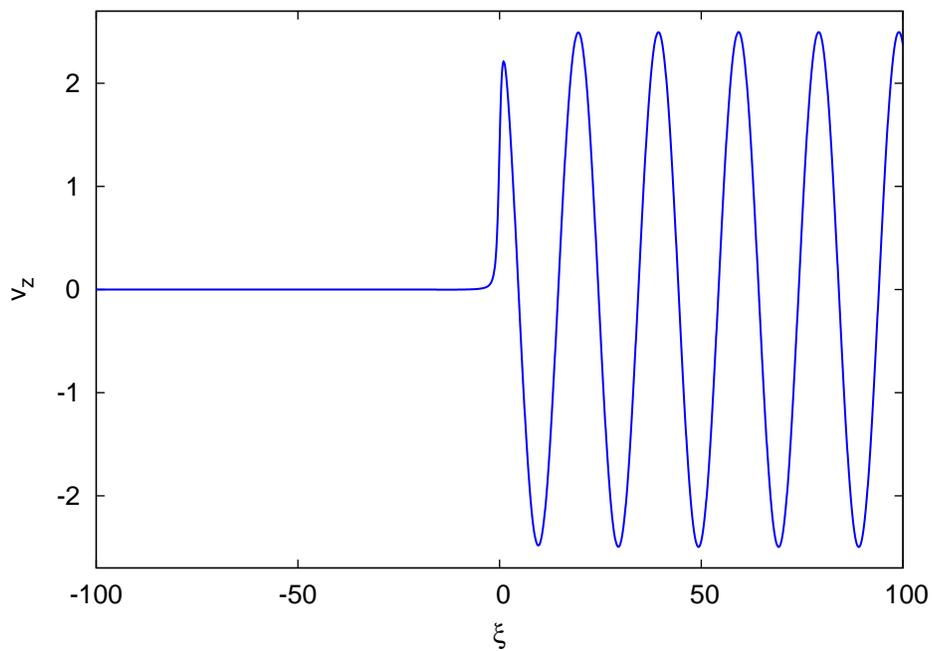} \caption{Out of plane velocity
oscillations parallel to the disk axis $\hat z$ as function of the
dimensionless time $v_z(\xi)$.
\label{fig:vz} }
\end{figure}
\begin{figure}
\centering\includegraphics{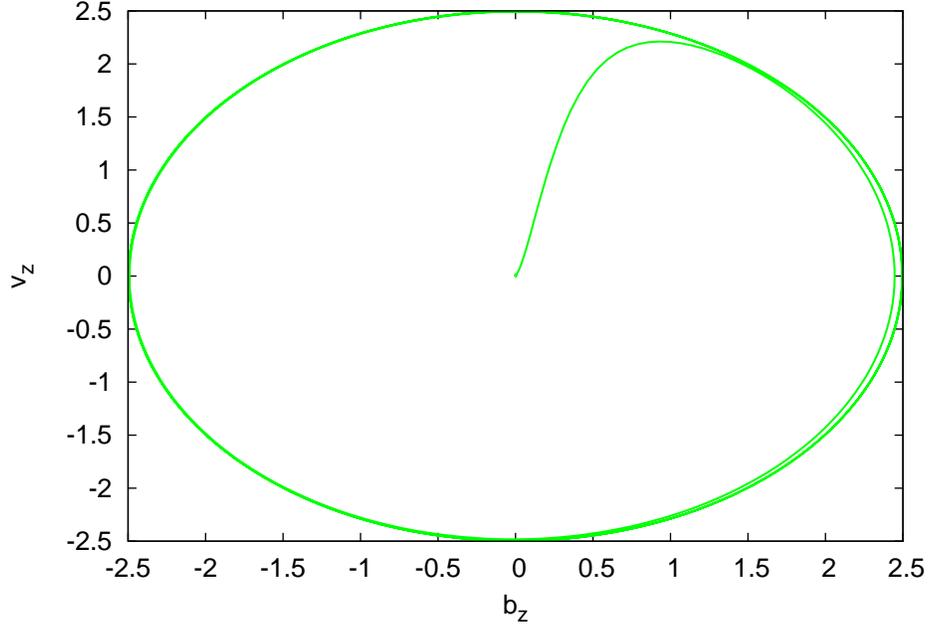} \caption{Phase portrait of
the AW oscillations transversal to the disk plane; $v_z$ versus
$b_z$. The AW polarization is almost perpendicular the constant
component of the magnetic field and the wavevector for $\xi=100.$
The wave amplitude at $\xi=-100$ is negligible; we have mode
conversion of SMW to AW at $\xi=0.$
\label{fig:bz_vs_vz} }
\end{figure}
\begin{figure}
\centering\includegraphics{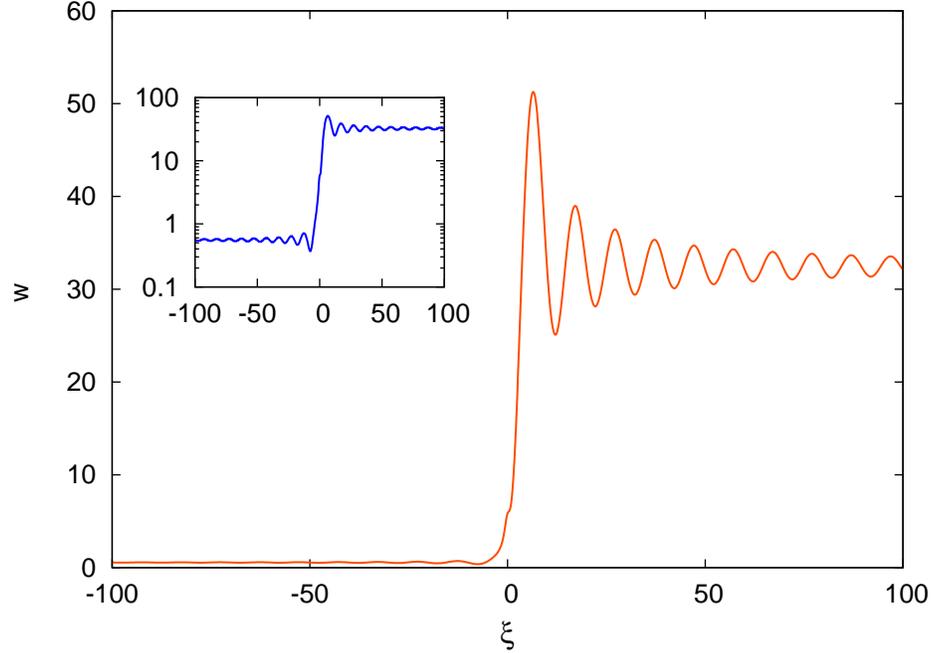} \caption{Dimensionles wave energy
density $w=\frac{1}{2}(\mathbf{v}^2+\mathbf{b}^2)$ versus
dimensionless time $\xi$. In the inset the ordinate is logarithmic.
The ratio $w(100)/w(-100)$ describes the energy amplification by the
shear flow which is the heating mechanism of accretion disks. The
analytical expression of the energy gain
Eqs.~(\ref{long_system_simpl}--\ref{long_system_last_simpl}) is
given by the confluent Heun functions Eqs.~(\ref{PsiHeun_g},
\ref{PsiHeun_u}). The approximative central symmetry of the blue
curve corresponds to $10$\% accuracy of the product $w(t)w(-t)\simeq
\mathrm{const}$ for this numerical example.
\label{fig:wx} }
\end{figure}
Those general formulas give a solution to the Cauchy problem. Having
in the beginning $t=t_0$ a distribution of the magnetic field
$\mathbf{B}_\mathrm{wave}(\mathrm{r},t_0)$ and velocity with
$\nabla\cdot\mathbf{V}_\mathrm{wave}(\mathrm{r},t_0)=0$, we can
calculate the Fourier components
\begin{eqnarray}
&&\mathbf{v}_\mathbf{k}(t_0)=\ii\int
\frac{\mathbf{V}_\mathrm{wave}(\mathrm{r},t_0)}{V_\mathrm{A}}
\mathrm{e}^{-\ii\mathbf{k}\cdot\mathbf{r}}\mathrm{d}x^3,\\
&&\mathbf{b}_\mathbf{k}(t_0)=\int
\frac{\mathbf{B}_\mathrm{wave}(\mathrm{r},t_0)}{B_0}
\mathrm{e}^{-\ii\mathbf{k}\cdot\mathbf{r}}\mathrm{d}x^3,\\
&&\xi_{0,\,\mathbf{k}}\equiv
-\frac{k_x}{\sqrt{k_y^2+k_z^2}},
\end{eqnarray}
and initial dimensionless time $\xi_{0,\,\mathbf{k}}$. If $k_z=0$
then $\mathrm{sgn}(k_y)\xi_{0,\,\mathbf{k}}=\tau_{0,\,\mathbf{k}}=-k_x/k_y$.  Then we
have to determine the coefficients $C$ in the general solutions for
$\psi$ \Eqref{gen} and $\chi$ \Eqref{chi_general} using the initial
values at $t_0$
\begin{eqnarray}
&&\mathbf{b}_\mathbf k(\xi)=C_\mathrm{g}\mathbf{b}_\mathrm{g}
          +C_\mathrm{u}\mathbf{b}_\mathrm{u}
          +\tilde{C}_\mathrm{g}\mathbf{b}_\mathrm{\tilde g}
          +\tilde{C}_\mathrm{u}\mathbf{b}_\mathrm{\tilde u},\\
&&\mathbf{v}_\mathbf k(\xi)=C_\mathrm{g}\mathbf{v}_\mathrm{g}
          +C_\mathrm{v}\mathbf{b}_\mathrm{u}
          +\tilde{C}_\mathrm{g}\mathbf{v}_\mathrm{\tilde g}
          +\tilde{C}_\mathrm{u}\mathbf{v}_\mathrm{\tilde u},\\
&&\mathbf{k}\cdot\mathbf{b}_\mathbf{k}
=0=\mathbf{k}\cdot\mathbf{v}_\mathbf{k}.
\end{eqnarray}
In this set we can use only $x$- and $z$-components, and so we
obtain $4$ equation for the constants $C_\mathrm{g}$,
$C_\mathrm{u}$, $\tilde{C}_\mathrm{g}$, and $\tilde{C}_\mathrm{u}$.
The functions $\mathbf{b}_\mathrm{g}(\xi)$ and
$\mathbf{v}_\mathrm{g}(\xi)$ are defined via substituting
$\psi_\mathrm g$ in
Eqs.~(\ref{long_system})--(\ref{long_system_last}) or
Eqs.~(\ref{long_system_simpl})--(\ref{long_system_last_simpl}), and
analogously $\psi_\mathrm u$, $\chi_\mathrm g$, and
$\chi_\mathrm{u}$.  Then at each moment $t$ we can calculate all
variables in the \textbf{k}-space
\begin{eqnarray}
&&\xi_\mathbf{k}(t)=\xi_{0,\,\mathbf{k}}
+(t-t_0)A\frac{k_y}{\sqrt{k_y^2+k_z^2}},\\
&&\mathbf{b}_\mathbf k(t)=\mathbf{b}_\mathbf k(\xi_\mathbf{k}(t)),\\
&&\mathbf{v}_\mathbf k(t)=\mathbf{v}_\mathbf k(\xi_\mathbf{k}(t)),\\
&&k_{x}^\mathrm{wave}(t)=k_x - (t-t_0)Ak_y.
\end{eqnarray}
Finally, we can return back to the real \textbf{r}-space
\begin{eqnarray}
&&\mathbf{V}_\mathrm{wave}(\mathrm{r},t)
=-\ii V_\mathrm{A} \!\int\!
\mathbf{v}_\mathbf{k}(t)
\,\mathrm{e}^{-\ii[\mathbf{k}\cdot\mathbf{r}-(t-t_0)Ak_yx]}
\,\frac{\mathrm{d}k^3}{(2\pi)^3},\nn\\
&&\mathbf{B}_\mathrm{wave}(\mathrm{r},t)
=B_0 \!\int\!
\mathbf{b}_\mathbf{k}(t)\,
\mathrm{e}^{-\ii[\mathbf{k}\cdot\mathbf{r}-(t-t_0)Ak_yx]}
\,\frac{\mathrm{d}k^3}{(2\pi)^3},\nn\\
&&\mathrm{Re}(\mathrm{e}^{-\ii\,\varphi})=
\cos \varphi,\qquad
\mathrm{Re}(-\ii\mathrm{e}^{-\ii\,\varphi})=-\sin \varphi.
\end{eqnarray}
This evolution of MHD variables is the main detail of the theory of
MHD turbulence in a shear flow.

Consider now an imaginary fluid filling the phase space ${\bf k}$
and $w(t) \equiv  \mathcal{E}_\mathbf{k}(t)$ from
\Eqref{density_of_rain} being the energy density carried by a
droplet of that fluid. As a wave mode initially with wave-vector
${\bf k}$ evolves according to \Eqref{q(t)}, the infinitesimal
phase-fluid droplet associated with that mode moves in the ${\bf
k}$-space. Wave amplification means that the energy density of the
droplets increases by a factor of $G$
\be
G=\frac{w(t\rightarrow+\infty)}{w(t\rightarrow-\infty)}.
\ee
Indeed for $\chi=0$, $k_z=0$, and $\xi\rightarrow\infty$
\be
b_y^2\asymp \psi^2\gg b_x^2+b_z^2,\qquad
v_y^2\asymp \left( \frac{\mathrm{d}_\xi\psi}{Q} \right)^2
\gg v_x^2+v_z^2
\ee
and
\be
w(t\rightarrow\infty)=\frac{1}{4}V_\mathrm{A}^2 D_\mathrm{f}^2,\qquad
w(t\rightarrow -\infty)=\frac{1}{4}V_\mathrm{A}^2.
\ee
For big enough time arguments $|\xi|\gg1$ and purely two-dimensional
waves with $k_z=0$ the motion of the fluid asymptotically
corresponds to a SMW with dispersion coinciding with the AW one
\begin{eqnarray}
&&Q\xi=\omega_{_\mathrm{SMW}}t,\qquad
\omega_{_\mathrm{SMW}}=\omega_{_\mathrm{AW}}= V_\mathrm{A}|k_y|,\nn\\
&&\psi(\xi)\asymp D_\mathrm{f}
\cos(\omega_{_\mathrm{SMW}}t+\phi_\mathrm{f}).
\end{eqnarray}
The Poynting vector, i.e., the energy flux in $\mathbf{r}$-space is
$V_\mathrm{A}w$.

The velocity of the droplet in the \textbf{k}-space according to
\Eqref{q(t)} determines the field of the shear flow in the
\textbf{k}-space
\be
\label{U_shear}
\mathbf{U}=\mathrm{d}_t\mathbf{q}(t)=-Aq_y\mathbf{e}_x,\qquad
\mathbf{U}^\mathrm{shear}_\mathbf{k}=-Ak_y\mathbf{e}_x.
\ee
Looking at a droplet we actually derive the shear flow velocity
field in \textbf{k}-space, $\mathbf{U}^\mathrm{shear}_\mathbf{k}$.

According to the Kolmogorov--Obukhov cascade of energy we have a
constant energy flux through each spherical surface with surface
element $\mathrm{d}\mathbf{f}$ in \textbf{k}-space
\be
\label{surface}
\varepsilon
=\oint
\mathcal{E}_\mathbf{k}^\mathrm{KO}\mathbf{U}^\mathrm{KO}_\mathbf{k}
\mathrm{d}\mathbf{f}
=\mathcal{E}_{k}^\mathrm{KO} {U}^\mathrm{KO}_{k}4\pi k^2,
\ee
which gives
\be
\mathbf{U}^\mathrm{KO}_\mathbf{k}\sim \varepsilon^{1/3}k^{5/3} \mathbf{e}_k,
\qquad
\mathbf{e}_k=\frac{\mathbf{k}}{k},
\ee
i.e., the velocity in \textbf{k}-space has dimension 1/(time$\times$length).
Here we used an important for our further work notion of the
energy flux in the \textbf{k}-space
\be
\mathbf{S}=\mathcal{E}_\mathrm{k}\mathbf{U}_\mathbf{k}
\ee
which is equal to energy density times velocity in the \textbf{k}-space.
This notion is analogous to the Poynting vector being, however,
defined in the \textbf{k}-space. In the Kolmogorov--Obukhov scenario we
have
\be
\frac{\partial}{\partial \mathbf{k}}\cdot \mathbf{S}^\mathrm{KO}
=\varepsilon
\delta(\mathbf{k}).
\ee

In order to approximate the turbulence as an initial source of MHD
waves we have to merge the turbulence with the wave spectral
densities and velocities. The simplest possible scenario is given in
the next subsection.

\subsection{Derivation of Shakura--Sunyaev Phenomenology in\\ %
the Framework of Kolmogorov Turbulence}                       %
%
How vortices create waves is a complicated problem far beyond the
scope of the present study. Here we will give only a model
illustration merging the spectral density of vortices
$\mathcal{E}^\mathrm{turb}_{\mathbf{k}}$ from Kolmogorov turbulence
with spectral density of magnetosonic waves
$\mathcal{E}^\mathrm{wave}_{\mathbf{k}}$
\be
\mathcal{E}^\mathrm{wave}_{\mathbf{k}}
\sim\mathcal{E}^\mathrm{turb}_{\mathbf{k}}
\sim \mathcal{E}_{_\Lambda}=
\varepsilon^{2/3}\Lambda^{11/3},
\ee
on the plane in momentum space
\be
k_x=-\mathrm{sgn}(k_y)\Lambda^{-1},\qquad
k_y^2+k_z^2<\Lambda^{-2},
\ee
where we qualitatively suppose that vortices are converted into
waves. Sign function corresponds to the direction of the shear flow
in the \textbf{k}-space, \Eqref{U_shear}. For $k_y$ we consider that
turbulent vortices have a given spectral density at
$k_x>\Lambda^{-1}$ which is converted to MHD wave energy at the
plane $k_x=\Lambda^{-1}$, and further on this wave energy evolves
according to our solution. In other words, the plane
$k_x>\Lambda^{-1}$ is the boundary between the vortex region and the
beginning of the amplification in the wave region where vortices
have negligible influence. In our qualitative picture we suppose
that vortices create spectral density which further on evolves as
wave spectral density with negligible influence.

The amplification \Eqref{gain} is essential $\mathcal G\gg1$ only within
a cylinder
\begin{equation}
\label{Gain_substitution}
\mathcal{G}(k_y, k_z) -1\sim \frac{1}{\Lambda^2 q^2},\qquad
q^2=k_y^2+k_z^2<\Lambda^{-2}
\end{equation}
with radius $\Lambda^{-1}$. This result with remains unchanged in
amplitude if we include the $J_\mathrm{c,u}$ and $J_\mathrm{s,g}$
terms.

The amplification occurs in the region
$-\Lambda^{-1}<k_x<\Lambda^{-1}$, that is to say from the cylinder
we cut a tube with length $2\Lambda^{-1}$.  In other words, we have a
domain with a shape of a tube in momentum space
\be
\mathcal{V}=\{
k_y^2+k_z^2<\Lambda^{-2},\qquad
|k_x|<\Lambda^{-1}\}.
\ee
In order to calculate the total power of waves $\mathcal{H}$ (per
unit mass) analogously to \Eqref{surface} we will integrate the
energy flux on the surface of the tube
\be
\label{tube}
\mathcal{H}=\oint
\mathcal{E}_\mathbf{k}^\mathrm{wave}\mathbf{U}^\mathrm{shear}_\mathbf{k}
\mathrm{d}\mathbf{f}.
\ee

As the shear in the physical flow results in a drift
of the wave modes along the axis of the tube,  we have to take into
account only the circular surfaces
\be
\epsilon\sim
\int
|U_x|\left[\mathcal{G}(k_y, k_z)-1\right]
\mathcal{E}_{_\Lambda}
\mathrm{d}k_y\mathrm{d}k_z.
\ee
The multiplier $(\mathcal G -1)$ takes into account the difference
between flowing out and flowing in energy fluxes.

We can use polar coordinates
\be
k_z=q \cos \theta,\qquad
k_y=q \sin \theta,\qquad
U_x=Aq \sin \theta.
\ee
Averaging over the angle $\theta$
\be
\langle U_x\rangle = \frac{2}{\pi}Aq
\sim A q,\;
\int_0^\pi\sin \theta \frac{\mathrm{d}\theta}{\pi}=\frac{2}{\pi},\;
\ee
and substituting it in \Eqref{tube}, using
$\mathrm{d}k_y\mathrm{d}k_z=\mathrm{d}(\pi q^2),$ leads to the
simple integral
\be
\mathcal{H}\sim
\int_0^{\Lambda^{-1}}
\frac{A q}{\Lambda^2 q^2}\,
\mathcal{E}_{_\Lambda}
q\mathrm{d}q
\sim \frac{A \mathcal{E}_{_\Lambda}}{\Lambda^{3}}\sim A V_\Lambda^2.
\ee
Then for the volume density of the amplified waves we have
\be
\mathcal{Q}\equiv\rho\mathcal{H}
\sim \rho A V_\Lambda^2
\sim \rho(\varepsilon V_A)^{2/3}A^{1/3}.
\ee
As all waves are finally dissipated, $\mathcal{Q}$
is actually the volume density of plasma heating.

For evanescent Kolmogorov turbulence power
\be
\rho \varepsilon \ll A\rho V_A^2=A B_0^2/\mu_0
\ee
the heating power $\mathcal{H}$ has a critical behavior
\be
\mathrm{d}_{\varepsilon}\mathcal{H}\sim
G_\mathrm{turb}\equiv\frac{\mathcal{H}}{\varepsilon}
\sim\left(\frac{AV_A^2}{\varepsilon}\right)^{1/3}\gg1,
\qquad
\mathcal{H}\gg\varepsilon
\qquad \mbox{for}\quad \varepsilon \rightarrow 0
\ee
which demonstrates that disks can ignite as a star even for very
weak turbulence and magnetic field. The ratio of wave power and
Kolmogorov vortex power $G_\mathrm{turb}$ can be considered as an
amplification coefficient for the turbulence. This energy gain shows
how efficient is the transformation of shear flow energy into waves
or in a broader framework the transformation of gravitational energy
into heat of accretion disks.

For hydrogen plasma $\rho c_\mathrm{s}^2/p=5/3\sim 1.$ Now we can
evaluate the shear stress (as given by the ratio of the volume
density of heating power and the shear frequency)
\be
\sigma=\frac{2\mathcal{Q}}{A}\sim
\rho \left(\frac{\varepsilon V_A}{A}\right)^{2/3}
\sim \rho V_\Lambda^2
\ee
via an effective viscosity
\be
\eta_\mathrm{eff}=\frac{\sigma}{A}\sim
\rho \frac{\left(\varepsilon V_A\right)^{2/3}}{A^{5/3}},\qquad
\nu_\mathrm{eff}=\frac{\eta_\mathrm{eff}}{\rho}
\sim\frac{\mathcal{H}}{A^2}
\sim \frac{\left(\varepsilon V_A\right)^{2/3}}{A^{5/3}}
\ee
and the dimensionless Shakura--Sunyaev friction coefficient
\be
\label{Shakura-Sunyaev}
\alpha\equiv \frac{\sigma}{p}\sim \frac{V_\Lambda^2}{c_\mathrm{s}^2}.
\ee
Including of the energy of $z$-polarized AWs does not modify this
result. Here we wish to emphasize that in our evaluation of the
energy gain, we were concentrated on the wave amplification of the
energy of two dimensional motion in the $x$--$y$ plane. Taking into
account the energy in $z$-direction shows that the heating is even
higher, which is of course in the favor of the concepts.

For an approximately  Keplerian disk rotation the shear rate is half
of the frequency of the orbital Keplerian angular velocity
$A=-\frac{1}{2}\omega_\mathrm{Kepler}$.  In this case for time
$A^{-1}$ the disk rotates per $2$ radians.  For Earth's rotation
along the Sun this time is of the order of one season. In such a way
the length parameter of our problem $\Lambda=V_\mathrm{A}/A$ can be
evaluated as one Alfv\'en season. Then $V_\Lambda$ from the final
result for the Shakura--Sunyaev parameter can be qualitatively
considered as a pulsation of the turbulent velocity for two disk
particles at distance equal to one Alfv\'en season $\Lambda$. Our
theory is formally applicable for $V_\Lambda\ll c_\mathrm{s}$ but
the boundary of its applicability (when compressibility effects stop
the SMWs amplification) allows us to understand that strong disk's
turbulence can lead to Shakura--Sunyaev upper limit $\alpha\sim1$.
Thus the following cascade of events emerges as a likely scenario
for the intense heating in accretion flows: the heating of the bulk
of the disk creates convection. For strong heating the convection is
turbulent. Turbulence generates magnetohydrodynamic waves. Waves are
amplified by the shear flow -- this is the transformation of
gravitational energy of orbiting plasma into waves. Waves finally are
absorbed by the viscosity which creates the heating. The heat is
emitted through the surface of the disk.  This process of formation
of stars and other compact astrophysical objects from nebulas works
continuously -- we have a self-consistent theory for self-sustained
turbulence of the magnetized accretion disks.

The weak point of this scenario is the supposed convective
turbulence which in presence of magnetic fields is unlikely to be of
Kolmogorov type. We consider as much more plausible scenario the
appearance of a self-sustained magneto-hydrodynamical turbulence
considered in the next subsection.

\subsection{Kraichnan Turbulence as a more Plausible Scenario for Accretion Disks}%
%
Magnetic field qualitatively changes the behavior of the fluid.  We
have no waves generated by vortices -- the turbulence in magnetic
field is related to MHD waves. Analogously to the Kolmogorov law
\Eqref{Kolmogorov}, for the Kraichnan turbulence the power of energy
cascade in the dissipation-free regime is given by the wave--wave
interaction
\be
\label{Kraichnan}
\varepsilon=
\frac{(V_\lambda^2)^2}{\lambda V_\mathrm{A}}
=\mathrm{\frac{(velocity)^3}{length}=\frac{power}{mass}}.
\ee
This power is proportional to the intensity of the two interacting waves
and this nonlinear effect for incompressible fluid is due to
the convective term $\mathbf{V}\cdot\nabla\mathbf{V}$ of the substantial
acceleration $\mathrm{D}_t\mathbf{V} = \partial_t\mathbf{V} +
\mathbf{V}\cdot\nabla\mathbf{V}$ of the momentum equation
\Eqref{MHD}.

The theory of generation of SMWs invokes parallels with the nonlinear
optical phenomena in lasers. The velocity oscillations of two amplified MHD waves
$\mathbf{V}^{(a)}$ and $\mathbf{V}^{(b)}$ create an external driving
force of the new wave with velocity field $\mathbf{V}$. In
the linearized \Eqref{MHD} we have to insert a small nonlinear
correction
\be
\rho \partial_t \mathbf{V} =
- \nabla p + \frac{\nabla\times\mathbf{B}}{\mu_0}\times \mathbf{B}
+\rho \mathbf{f},
\qquad \mathbf{f}\equiv \frac{1}{2}\sum_{a,\,b}
\mathbf{V}^{(a)}\cdot\nabla\mathbf{V}^{(b)}.
\ee
Here, in the inhomogeneous term $\mathbf{f}$ we have to perform
summation over all other MHD waves. This external for the wave force
(per unit mass) acts as an external noise and its statistical
properties are determined by the force--force correlator
\be
\hat\Gamma(\mathbf{r}_1,t_1;\mathbf{r}_2,t_2)
=\langle \mathbf{f}(\mathbf{r}_1,t_1)\,\mathbf{f}(\mathbf{r}_2,t_2) \rangle,
\ee
where the averaging is over the waves phases. A scenario of such
type (a Langevin MHD) was described in
Ref.~[\onlinecite{Mishonov:07}]; this approach is similar in the
spirit to the forced burgers turbulence.\cite{Woyczynski98} In the
framework of that scenario the strongly amplified
$|D_\mathrm{f}|\gg1$ MHD waves with asymptotics \Eqref{wave-asymp}
\be
\label{wave-asymp-damping}
\psi \approx
D_\mathrm{f}\,\theta(\xi)\cos(Q\xi+\phi_\mathrm{f})
\exp \left(-\nu^\prime K_y^2 \tau^3/6\right)
\ee
generate new waves and after a statistical averaging we have a
self-consistent theory for magnetic turbulence in a shear flow. So
MHD waves ignite the chain reaction of quasar self-heating. The last
exponential term describes the wave damping when a small viscosity
is taken into account. Damping is significant only for
$t\rightarrow\infty$ when $|k_x|\gg |k_y|$ and the wave-vector is
almost parallel to the magnetic field. In this geometry, the damping
rate of the wave density of AWs and SMWs is proportional to the
square of the frequency\cite{LL8}
\be
w(t)=w(0)\exp \left( -\frac{\omega^2}{V_\mathrm{A}^2}\nu t \right)
=w(0)\exp \left( -\nu k^2 t \right).
\ee
For the time-dependent wave-vector $k^2(t)\approx (k_y At)^2$ in the
argument of the exponent we have to make the replacement
\be
\nu k^2 t \rightarrow \nu \int_0^t k^2(t') \mathrm{d}t'
=\frac{1}{3}\nu^\prime K_y^2 \tau^3,\qquad \nu'
=\frac{\nu A}{V_\mathrm{A}^2}=\frac{1}{\mathcal{R}},
\qquad\mathcal{R}\equiv\frac{\Lambda V_\mathrm{A}}{\nu_\mathrm{k}},
\qquad\mathcal{S}_\mathrm{MRI}\equiv\frac{\Lambda V_\mathrm{A}}{\nu_\mathrm{m}},
\ee
where $\nu'$ is the dimensionless viscosity and
$\mathcal{R}=V_\mathrm{A}^2/A$ is the ``Reynolds number of the
magnetorotational instability (MRI),\cite{Masada:08}'' and
$\mathcal{S}_\mathrm{MRI}$ is the Lundquist number of MRI. After
long enough time $t_\nu$ when
\be
|k_x(t_\nu)|=1/\lambda_\mathrm{a},
\qquad
\lambda_\mathrm{a}=\frac{\nu}{V_\mathrm A},
\ee
MHD waves are completely dissipated.

The details of self-consistent MHD turbulence will be given
elsewhere, but again the wave amplification operates as a turbulence
amplifier. For MHD turbulence one can expect
\be
\label{tusj5}
\sigma_{R\varphi}=\alpha_\mathrm{m}(\nu') p_B,
\qquad p_B=\frac{1}{2}\rho V_\mathrm{A}^2. \ee
The evaluation of the magnetic friction coefficient $\alpha_\mathrm{m}$
as a function of the dimensionless viscosity is a new problem addressed
to the theoretical astrophysics.

\section{Discussions, Conclusions, and Perspectives} %
%
We propose that the long-missing element of the dissipation
mechanism of accretion disks is now identified -- it is the
amplification of long-wavelength SMWs. This is the indispensable
ingredient of the stars generating engine. Without it the Universe
would possibly be a structureless gas -- deserted and uninhabited.
The two-dimensional disks and the redistribution of angular momentum
in them are the means of creating a one-dimensional compact
astrophysical object from gases and dust. The planetary system is
the result of this  star-producing sequence of events when the
accretion stops and the disk is frozen. It is remarkable that a
simple equation of the Schr\"odinger type, \Eqref{Sch}, is at the
core of the friction mechanism which cheated the diversity of the
Universe. This means that the Schr\"odinger equation can describe
one more phenomenon while science is on its way to explaining the
frogs and the musical composers. ``The next great era of awakening
of human intellect may well produce a method of understanding the
qualitative content of equations. Today we cannot. Today we cannot
see that the water flow equations contain such things as the barber
pole structure of turbulence that one sees between rotating
cylinders. Today we cannot see whether Schr\"odinger's equation
contains frogs, musical composers, or morality -- or whether it does
not.\cite{Feynman:64}''

Notably, here we have observed only an amplification in a weak
magnetic field but not instability, where the amplitude of waves
increases infinitely with time. Also, the rotation is found
not relevant for this phenomenon. Therefore, regardless of similarity
in spirit, our work is
completely different from the research focused on rotational
instabilities in strong magnetic fields. Our results point to an amplification
but not to an instability. For sufficiently long  wavelengths the
amplification can be enormous but never infinite. Amplification means
exponential increasing of the amplitude in the framework of linear
theory.

After investigating the local dissipation and shear tension by the
self-consistent statistical MHD method the corresponding numerical
value of the $\alpha$ parameter can be incorporated in global models
for accretion disks dynamics. The global models for accretion disks
include also the problem of disk dynamo.\cite{Guenter:04} In order
to create a self-consistent magnetic field $B_{0,\varphi}$ the
accretion disks operate as radial inflow generators from plasma
physics,\cite{Boyd:69} see also Ref.~\onlinecite{Yoshizawa:02}. In
this broad program the local and detailed investigation of
magnetosonic waves propagation in a homogeneous shear flow and
magnetic field is only the first step in our understanding of the
properties of space plasmas, i.e., the low-gradient approximation is
the indispensable step in our understanding of accretion power in
the Universe.\cite{Juhan:02}

The further development of the theory of MHD waves turbulence will
give additional important details but even with what we know now, we
get an insight into the workings of the star-creating engine. A very
powerful, hence possibly dominant, mechanism of energy
transformation in shear magnetohydrodynamic flows is identified,
which is the amplification of Alfv\'en waves. They are also
responsible for the heating of the solar corona\cite{Stasiewicz:06}
and represent a quite common mechanism for heating of space plasmas
in general. Here we wish to recall that shear flows are important
for formation of toroidal magnetic field from poloidal one. The
differential rotation\cite{Ashwanden:06} of the Sun is another shear
flow which can lead to amplification of MHD waves. The convection
excites MHD waves in almost toroidal magnetic fluxtubes. When
buoyant magnetic fluxtubes reach photospheric surface the MHD waves
can be significantly amplified.

Recently Alfv\'en waves of sufficient strength have been
unambiguously observed in the chromosphere by Solar Optical
Telescope onboard the Japanese \emph{Hinode\/} satellite. Such
Alfv\'en waves are energetic enough to accelerate the solar wind and
to heat the quiet corona.\cite{DePontieu:07} In order to reach
quantitative agreement of the theory it is necessary to merge the
spectral density $\propto D_\mathrm{AW}/\omega^n$ of these AW  with
the speed of the solar wind $v_\mathrm{wind}$ and coronal
temperature $T_p.$ The mission of the theoretical models is to give
the simple relations between those experimentally accessible
parameters $f(T_p,v_\mathrm{wind},D_\mathrm{AW})=0,$ cf.
Ref.~[\onlinecite{Mishonov:07a}]. Heating of tokamak plasmas by
magnetosonic waves is also a widely discussed
issue.\cite{Mazurenko:01,Elfimov:02} According our scenario the
self-heated accretion discs are just working tokamaks.

The idea of accretion disks can be traced in the development of
contemporary science.\cite{Kippenhahn:87} While for Descartes
(1644), Kant, and Laplace that idea was a creative
mythology,\cite{Campbell} the Hubble Space Telescope has now
delivered observational evidence for disks of ionized gases around
massive black holes.\cite{Ford:94} Observations of  protoplanetary
disks are discussed at the web-page\cite{proplyd} and its
links.\cite{Mamajek:04,White:05,Wang:06}


It was only several decades after a horrific punishment for the
expression of now mainstream views took place at the market place
Campo dei Fiori\cite{Bruno:17Feb1600} in Rome, when Galileo observed
spots on the Sun and detected the Sun's rotation. Why the Sun is
rotating so slowly (i.e., the problem of angular momentum
dissipation) eventually becomes a focus issue in
cosmogony.\cite{cosmogony} The development of plasma physics
determined that the molecular viscosity of plasma is far too small
to account for the observed heating and that new ideas should be
tested. The amplification of SMWs, an idea with the potential of
explaining the enormous luminosity of quasars, was explored in the
present work.

Pursuing further applications and proof of concepts, one could
notice that the magnetosonic waves of magnetic turbulence can emit
radio waves through the large surface of the accretion disk.
Correlation between the radio waves and optical emission from
quasars will be the crucial test for the present theory of dissipation
in accretion disks.

\section{Acknowledgments}
The authors thank Nikolai~Shakura, Dmitry~Bisikalo, Temur
Zakarashvili, and Grigol Gogoberidze for the interest to the work
and Professor~Ivan~Zhelyazkov for critical reading of the manuscript
and many creative suggestions. This work was partially supported by
University of Sofia scientific grant from May 2009.

%

%
%
\end{document}